\documentclass{article}

\usepackage[a4paper]{geometry}
\usepackage[utf8]{inputenc}
\usepackage{graphicx}
\usepackage{latexsym}

\usepackage{lipsum}
\usepackage{graphicx}
\usepackage{authblk}
\usepackage[english]{babel}
\usepackage{color}
\usepackage{hyperref}
\hypersetup{
    colorlinks=true,
    linkcolor=blue,
    filecolor=blue,      
    urlcolor=blue,
    citecolor=blue,
    }
\usepackage{dsfont}
\usepackage{geometry}
\usepackage{float}
\usepackage{bbm}

\usepackage[nottoc]{tocbibind}
 \usepackage{indentfirst}

\graphicspath{{./figure/}}

\usepackage{pgfplots}
\usepackage{comment}

\usepackage{tikz}
\usepackage{caption}
\usepackage{subcaption}
\DeclareCaptionLabelFormat{custom}
{%
      \textbf{Figure #2}
}
\DeclareCaptionLabelSeparator{custom}{ }
\DeclareCaptionFormat{custom}
{%
\centering
    #1 #2 #3 
}
\usepackage{amsmath,amsfonts,amssymb,amsthm}

\numberwithin{equation}{section}
\usepackage{etoolbox}
\apptocmd{\thebibliography}{}{}{}
\newcommand{\R}{{\mathbb R}}

\theoremstyle{plain}

\newtheorem{reminder*}{[theorem]Reminder}

\newtheorem{details*}[theorem]{Details}

\newtheorem{comm*}{Comment}
 
\newtheorem{definition*}{[theorem]Definition}

\newtheorem{notation*}{Notation}

\date{}
\title{Coagulation equations with source leading to anomalous self-similarity}
\author{M. A. Ferreira, E. Franco, J. Lukkarinen, A. Nota, J. J. L. Vel\'{a}zquez}

\begin{document}
\maketitle
\begin{abstract}
We study the long-time behaviour of the solutions to Smoluchowski coagulation equations with a source term 
%a class of coagulation equations including a source term
of small clusters. The source drives the system out-of-equilibrium,  leading to a rich range of different possible long-time behaviours, including anomalous self-similarity.
The coagulation kernel is non-gelling, homogeneous, with homogeneity $\gamma \leq -1 $, and behaves like $x^{\gamma+\lambda} y^{-\lambda} $ when $y \ll x$ with $\gamma+2\lambda > 1 $. %\mf{ Remove? We also review results obtained in the complementary regimes for the parameters $\gamma$ and $\lambda$.} 
Our analysis shows that the long-time behaviour of the solutions  depends on the parameters $\gamma$ and $\lambda$. More precisely, we argue that the long-time behaviour is self-similar, although the scaling of the self-similar solutions depends on the sign of $\gamma+\lambda$ and on whether $\gamma=-1$ or $\gamma < -1$. 
In all these cases, the scaling differs from the usual one 
%of self-similar solutions
that has been previously obtained when $\gamma+2\lambda <1$ or $\gamma+2\lambda \geq 1, \gamma >-1$. 
In the last part of the paper, we present some conjectures supporting the self-similar ansatz also for the critical case $\gamma+2\lambda=1, \gamma \leq -1 $.
\end{abstract}
\bigskip 

\textbf{Keywords:} Smoluchowski coagulation equations; injection; non-equilibrium; long-time behaviour; dimensional analysis

\tableofcontents

\section{Introduction}

\subsection{Motivation}

In this paper, we study the long-time behaviour of solutions to
the Smoluchowski coagulation equation with source (or injection).
A commonly used approach is to assume that the concentration distribution, after a proper choice of a  scaling function $L(t)$, asymptotically approaches
a stationary shape: the shape function is then called a {\em self-similar profile}  to the coagulation equation.
Some  references about successful uses of self-similar solutions to Smoluchowski coagulation equations in the mathematical and physical literature can be found here \cite{aldous1999deterministic, banasiak2019analytic,chandrasekhar1943stochastic, Friedlander, Smoluchowski}.

Once this  dynamical scaling hypothesis, i.e.,  the existence of a self-similar scale, has been made, mere dimensional considerations
allow to determine the time-dependence of the scale for a large class of homogenenous coagulation rate kernels $K(x,y)$, which behave like $x^{\gamma+\lambda}y^{-\lambda}+y^{\gamma+\lambda}x^{-\lambda}$ for some parameters $\gamma, \lambda \in \R$.  

 In the absence of gelation, it has been  shown that the scaling  $L(t) := t^{\frac{1+s}{1-\gamma}}$ allows to find self-similar solutions to the coagulation equation, both in the case without source ($s=0$) and in the case with a constant source ($s=1$), provided that  the condition $\gamma+2 \lambda<1$ holds in the latter case. We will call such solutions {\em standard self-similar solutions}.

%In terms of the standard homogenity parameters $\gamma, \lambda$ of the rate kernel, it has been shown that, if $\gamma+2 \lambda<1$, then scaling with $L(t) = t^{\frac{1+s}{1-\gamma}}$ allows to find self-similar solutions to the coagulation equation. 
%  If there is no source, here $s=0$ (cf.\ Sec.\ \ref{sec:dimensional without source}), and if there is a constant source, then $s=1$ (cf.\ Sec.\ \ref{sec:standardselfsim}).
%We will call such solutions {\em standard self-similar solutions}.

We emphasize that the presence of a source term can strongly affect the long-time behaviour. 
 Indeed, in the regime 
%It turns out that, if we relax the above assumption and consider cases which have a source and
$\gamma+2 \lambda\ge 1$,  self-similar solutions can still exist but one needs to allow for more complex dependence of the scaling on the parameters of the rate kernel.  In this work, we first  summarize known  results about the long-time  properties of solutions to the Smoluchowski coagulation equation with a constant source term, including stationary solutions when they exist. 
Then we study new self-similar solutions which exhibit {\em anomalous scaling}.
These include cases where $L(t) \neq t^{\frac{1+s}{1-\gamma}}$ or where more detailed analysis is needed to capture the asymptotic form correctly. Our aim in this paper  is to provide plausible conjectures about the long-time behaviour using formal asymptotics. 

%\eu{Remove? " We do not aim for mathematical rigour, but rather to give plausible conjectures based on previous works."}

The physical motivation to look for such solutions to coagulation equations with a source comes from two particular examples: from
coagulation processes relevant in atmospheric science, see for instance \cite{Friedlander}, \cite{Vehkamaki+others} and \cite{Vehkamaki}, and from applications in epitaxial growth, see \cite{KMR98}, \cite{KMR99}.
In the first case, the commonly used free molecular rate kernel belongs to the regime where $\gamma+2 \lambda> 1$, and we recall in Sec.\ \ref{sec:ballisticselfsim} the results from  \cite{cristian2022long} about how the standard self-similar ansatz needs to be adjusted in this case.   For epitaxial growth, a wide range of values of $\gamma$ and $\lambda$ might occur, including some of the anomalous marginal cases discussed here.
The physical motivation and relevant parameter values in epitaxial growth are discussed, for example, in
\cite[Sec. II]{KMR99}.

\subsection{Discrete and continuous coagulation equation with source}

We are interested in
the behaviour of the solutions to the following  spatially homogeneous problem
\begin{equation}
\partial_{t}f\left(  t, x\right)  =\mathbb{K}\left[  f\right]  \left(t,
x\right)  +\eta\left(  x \right)  \label{S1E1}%
\end{equation}
where $\mathbb{K}\left[  f\right]  $ is the classical Smoluchowski coagulation
operator%
\begin{equation}
\mathbb{K}\left[  f\right]  \left( t, x\right)  =\frac{1}{2}\int_{0}%
^{x}K\left(  x-y,y\right)  f\left(  t, x-y\right)  f\left(t,  y\right)
dy-\int_{0}^{\infty}K\left(  x,y\right)  f\left(  t, x\right)  f\left(
t,y\right)  dy \,. \label{S1E2}%
\end{equation}
Here, $K\left( x,y\right)  $ is the coagulation rate kernel whose form
depends on the specific microscopic mechanism resulting in the coagulation of the monomer clusters.
We will assume in the following that $K\left(  x,y\right)  =K\left(
y,x\right)  .$ The function $f\left(  t, x\right)  $ denotes the distribution function for clusters with size $x$ at time $t$. The function
$\eta\left(  x\right)  $ in (\ref{S1E1}) is a source of clusters with volume
$x$ which we take to be constant in time. We will assume that $\eta\left(  x\right)  $ decreases fast enough for
large sizes $x$.  To be concrete, most of the mathematical results will assume that $\eta\left(  x\right)  $ is supported in the interval $[1, L_\eta] $; this sets the scale $x=1$ to correspond to monomers and $x=L_\eta$ is an upper bound for the size of injected clusters.
As shown, for instance in \cite{dubovskiui1994mathematical} and in \cite{ferreira2019stationary}, but also in Section \ref{sec:dimensional}, the presence of the source in equation \eqref{S1E1} enriches the dynamics of the coagulation equation and leads to intricate, mathematically challenging problems.

In many applications of the Smoluchowski coagulation equations, the sizes of
the clusters described by the equation take only a set of discrete values
because all the clusters are just aggregates of  monomers. Assuming that the
size of each monomer is one, we can obtain the discrete coagulation equations by
looking for solutions to (\ref{S1E1}), (\ref{S1E2}) with the form%
\begin{equation}
f\left(  t, x\right)  =\sum_{k=1}^{\infty}n_{k}\left(  t\right)  \delta\left(
x-k\right)  \label{S1E3}%
\end{equation}
where%
\begin{equation}
\eta\left(  x\right)  =\sum_{k=1}^{\infty}\eta_{k}\delta\left(  x-k\right)
\label{S1E4}%
\end{equation}
Then (\ref{S1E1}), (\ref{S1E2}) reduces to the following system of infinitely
many ordinary differential equations%
\begin{equation}
\partial_{t}n_{k}=\frac{1}{2}\sum_{\ell=1}^{k-1}K\left(  k-\ell,\ell\right)
n_{k-\ell}n_{\ell}-\sum_{\ell=1}^{\infty}K\left(  k,\ell\right)  n_{k}n_{\ell
}+\eta_{k}. \label{S1E5}%
\end{equation}
We will here focus on the continuous case, keeping in mind that often one can use the continuous self-similar solutions also to approximate large-time evolution of similarly scaled solutions to the discrete equation
\cite{eichenberg2021self,ferreira2019stationary}.

In this work, we assume that the coagulation kernel is \textit{homogeneous}, i.e., there exists a $\gamma \in \mathbb R$ such that
\begin{equation}
K\left(  ax,ay\right)  =a^{\gamma}K\left(  x,y\right)
\ \ ,\ x,y>0,\ a>0. \label{S1E6}%
\end{equation}%
Then, the kernel $K$ can be written in the following way
\begin{equation}
K\left(  x,y\right)  =\left(  x+y\right)  ^{\gamma}F\left(  \frac{x}%
{x+y}\right)  \ \text{with\ }F \left(  s\right)  =F\left(  1-s\right)
\ \ ,\ s\in\left(  0,1\right),  \label{S1E7}%
\end{equation}
where the function $F$ is continuous.

We additionally assume that $K(x,y) \approx y^{\alpha} x^{\beta} $ when $ x \ll y$ and $K(x,y) \approx x^{\alpha} y^{\beta} $ when $ y \ll x$, for some $\alpha, \beta \in \mathbb R.$
This assumption on the behaviour of the coagulation rate between clusters of very different sizes is equivalent to assuming the following asymptotic behaviour for the function $F$ near the origin
\begin{equation}
F \left(  s\right)  \sim\frac{1}{s^{\lambda}}\ \ \text{as\ \ }
s\rightarrow0^{+}\label{S1E8}
\end{equation}
where, without loss of generality, we are taking $\gamma+\lambda= \max{\{\alpha, \beta \}}$ and $\lambda=-\min{\{\alpha, \beta\} }$, hence $\gamma+2\lambda\geq0$.

We also restrict our attention here to the case in which \textit{gelation} does not take place.
To this end, we make the following assumption on the parameters $\gamma$ and $\lambda $
\begin{equation} \label{non-gelling assumption}
 \gamma < 1 \text{ and } \gamma + \lambda < 1.
\end{equation}
We refer to \cite{davies1999smoluchowski} for more details about the gelation regimes for some ranges of exponents.
Since we are considering a coagulation equation with a source term, the non-gelation assumption does not imply mass conservation but, instead, it corresponds to a linear increase of the total mass in time.
This is crucial for the dimensional considerations done in Section \ref{sec:dimensional}. More precisely,
as we will see in Section \ref{sec:dimensional}, different assumptions on the parameters $\gamma$ and $\lambda $  lead to different behaviour of some of the moments of the solution $f$ to equation \eqref{S1E1}.
It is therefore useful to introduce the following notation:
\begin{equation} \label{C6}%
M_{\mu}=\int_{0}^{\infty} x ^{\mu
}f\left(  t,x\right)  dx, \ \mu \in \mathbb R.
\end{equation}
We will also consider, see Section \ref{sec:inner region gamma+lambda to infinity}, the moments of the discrete equation \eqref{S1E5}. These will be denoted as
\begin{equation}
m_{J}=\sum_{k=1}^{\infty} k ^{J}c_{k}\ \ \text{for each }%
J\in\mathbb{R}. \label{A5}%
\end{equation}
Two of these moments will play a central role in determining the long-time
asymptotics of the cluster distribution of the system, namely, $M_{0}$ and $M_{1}$ which
represent, respectively, the total number of clusters of the system and the total
number of monomers contained in all of the clusters.
%\begin{equation}
%M_{\gamma+\lambda}=\int_{0}^{\infty}\left(  x\right)  ^{\gamma+\lambda
%}f\left(  x,t\right)  dx.\label{C6}%
%\end{equation}

The assumptions on the kernel $K$ that we make in this paper,
 are slightly stronger than the assumptions made in \cite{ferreira2019stationary}, \cite{ferreira2021self} and \cite{cristian2022long}. In these papers, it was enough to assume upper and lower bounds of the form $x^{\gamma+\lambda} y^{-\lambda} +y^{\gamma+\lambda} x^{-\lambda} $ for the kernel $K$, while here we need to prescribe the asymptotic behaviour as $x$, $y$ tend to zero or to infinity.
 These assumptions will be needed to compute the detailed asymptotics of the solutions $f$ to equation \eqref{S1E1}. 
% The interest in these class of kernels was initially motivated by the kernels in atmospheric science, for instance the free molecular kernel and the Brownian kernel.

\section{Classification of the behaviour for different parameter regimes}
\subsection{Review of previous  results} \label{sec:ExisResults}

\subsubsection*{Existence/non-existence of stationary solutions: the role of the critical parameter $\gamma+2\lambda =1$ }

Since we are considering a coagulation equation with source, one interesting problem is to study the existence of non-equilibrium stationary solutions for equation \eqref{S1E1}, i.e., solutions to the following equation
\begin{equation} \label{eq:stationary}
\mathbb K f = \eta .
\end{equation}
For the class of kernels we are interested in, the existence of solutions to \eqref{eq:stationary} has been proven in \cite{ferreira2019stationary} under the additional assumption that $\gamma+2 \lambda <1 $.
Moreover, the solutions $f$ to \eqref{eq:stationary} behave as the power law $x^{- \frac{\gamma+3}{2}}$ for large values of $x$.  
In \cite{ferreira2019stationary}, it is also proven that solutions to \eqref{eq:stationary} do not exist if $\gamma+2\lambda \geq 1$.
These results are in agreement with the formal asymptotics obtained in \cite{hayakawa1987irreversible} for the discrete equation \eqref{S1E5}.

The argument used  in \cite{ferreira2019stationary} to prove the existence of solutions to \eqref{eq:stationary} when $\gamma+2\lambda <1$ is based on the fact that the main contribution to the flux of mass, from the source to the region of large clusters, is given by the interaction of clusters of similar sizes, while the contribution due to the interaction of clusters of very different sizes is negligible.
In contrast, the argument to prove non-existence when $\gamma+2\lambda \geq 1 $ is based on the fact that, for this range of exponents, the transfer of clusters of size $x$ of order one towards very large cluster sizes is so fast that the concentration of clusters with size of order one would become zero at stationarity, hence a contradiction.
It is therefore  natural to expect that $\gamma+2\lambda=1 $ is a critical value and to expect different behaviours for the solutions to equation \eqref{S1E1} when $\gamma+2 \lambda<1$ and $\gamma+2\lambda \geq 1$.

As explained in \cite{ferreira2021self}, for the case $\gamma+2\lambda <1$, and in \cite{cristian2022long}, for $\gamma+2\lambda>1, \gamma >-1$, the solutions to equation \eqref{eq:stationary} are of great interest for the analysis of the long-time behaviour of the solutions to equation \eqref{S1E1}. 
Indeed, they describe the behaviour of the solutions $f$ to equation \eqref{S1E1} in the region of small sizes.

The results in \cite{ferreira2019stationary} suggest that when $\gamma+2\lambda < 1$ then $f(t, x) \sim x^{-\frac{\gamma+3}{2}}$ as $t \rightarrow \infty $ and when $x $ is of order $1$. 
Instead, when $\gamma+2\lambda >1$ and $\gamma > -1$, the analysis in \cite{ferreira2019stationary} suggests that $f(t, x) \rightarrow 0$ as $t \rightarrow \infty $ and when $x $ is of order $1$. 
From these considerations we clearly see that $\gamma + 2\lambda=1 $ is a threshold case at the intersection of different behaviours for the solution $f$ to equation \eqref{S1E1}.

\subsubsection*{Standard self-similar long-time behaviour when $\gamma+2\lambda < 1 $}\label{sec:standardselfsim}

In \cite{ferreira2021self} the long-time behaviour of the solutions to equation \eqref{S1E1} is analysed when $\gamma+2\lambda < 1$.
In that case, the scaling properties of the kernel $K$, together with the existence of a solution to equation \eqref{eq:stationary} lead to the following self-similar ansatz with standard scale $L(t)$: the solutions $f$ to equation \eqref{S1E1} behave, for large times and large sizes, as
\begin{equation} \label{standard self similar solution}
f(t,x)= %\frac{1}{t^{\frac{\gamma+3}{1-\gamma}}}
\frac{t}{L(t)^2} \Phi(\xi) \ \text{ with } \ \xi= \frac{x}{L(t)}\,, \quad L(t) := t^{\frac{2}{1-\gamma}}
\end{equation}
where the self-similar profile $\Phi$ satisfies the following equation
\begin{equation}
-\frac{2}{1-\gamma}\xi\Phi_{\xi}-\frac{3+\gamma}{1-\gamma}\Phi=\mathbb{K}
\left[  \Phi\right], \label{SelfSimFlux}
\end{equation}
with the boundary condition
\begin{equation}
\lim_{\xi\rightarrow0}\int_{0}^{\xi}\int_{\xi-x}^\infty K\left( x
,y\right) x\Phi\left( x\right)  \Phi\left(  y \right)  dy dx = \int_{0}^{\infty}x\eta\left(  x\right) d x.
\label{SelfSimFlBC}
\end{equation}
This ansatz is also supported by the numerical study conducted in \cite{davies1999smoluchowski}. The fact that we are considering a coagulation equation with a source term is, in this case, reflected in the boundary condition \eqref{SelfSimFlBC} which is absent in the theory of self similar solutions to the classical coagulation equation.

\subsubsection*{Standard self-similar long-time behaviour when $\gamma+2\lambda >1$ and $\gamma >-1$}
\label{sec:ballisticselfsim}

The long-term behaviour in the complementary case $\gamma+2\lambda \geq 1 $, has been studied in \cite{cristian2022long} for $\gamma>-1$.
As explained in that paper, we expect two slightly different behaviours to occur when $\gamma+2\lambda > 1 $ and when $\gamma+2\lambda =1$.
Indeed, dimensional arguments that will be presented in detail in Section \ref{sec:dimensional}, imply that if $\gamma+2\lambda >1$ and if we assume standard self-similar scalings, then the moment $M_{\gamma+\lambda} \rightarrow \infty $ as time goes to infinity, while if $\gamma+2\lambda =1 $, then $M_{\gamma+\lambda}$ converges to a constant as time goes to infinity. For this reason, the case $\gamma+2\lambda =1 $ will be separately  mentioned in Section \ref{sec:overview}. 

When we assume the kernel $K$ to be as in \eqref{S1E6}, \eqref{S1E7}, the loss term in the coagulation operator \eqref{S1E2} depends on the moment $M_{\gamma+\lambda}$ and tends to infinity if $M_{\gamma+\lambda}$ tends to infinity. Therefore, when $\gamma+2\lambda >1 $, we expect, in the region of small clusters, the loss term of the coagulation operator to tend to infinity and the solutions $f$ to equation \eqref{S1E1} to tend to zero as time goes to infinity. Notice that this scenario is compatible with the non-existence of stationary solutions.

In contrast, in the region of large cluster sizes, we can approximate part of the coagulation operator in equation \eqref{S1E1} with a transport term, representing the coagulation of the small clusters injected by the source with very large clusters. Hence, we expect the solutions $f$ to equation \eqref{S1E1} to behave like the solutions to the following equation for large sizes and large times
    \begin{equation} \label{eq:coag with growth}
        \partial_t f(t,x) + \partial_x \left( \frac{x^{\gamma+\lambda}}{M_{\gamma+\lambda}} f(t,x) \right)  = \mathbb K f(t,x).
    \end{equation}
    A heuristic derivation of this equation is presented in Section \ref{sec:regions}.
Notice that equation \eqref{eq:coag with growth} is compatible with the existence of self-similar solutions, i.e., solutions that are invariant under the rescalings that preserve the form of the equation and that are compatible with the linear growth of the mass, i.e., solutions of the form \eqref{standard self similar solution}.

It is proven in \cite{cristian2022long} that, when $\gamma+ 2 \lambda >1$, self-similar solutions of the form \eqref{standard self similar solution} exist if and only if $\gamma > -1$.  The profile equation (\ref{SelfSimFlux}) merely attains a new term corresponding to the additional growth term in (\ref{eq:coag with growth}).
Some properties of the self-similar solutions in the case $\gamma+2\lambda > 1 $ and $\gamma > -1 $ have been studied in \cite{cristian2022long}. In particular, it is proven there that the self-similar solutions are zero near zero and decay exponentially for large sizes.

The reason why the assumption $\gamma \leq -1 $ is not compatible with the existence of self-similar solutions of the form \eqref{standard self similar solution} is that, in that case, the $0$-th moment of the self-similar solution, $M_0(f_s)$, grows in time as $ t^{- \frac{\gamma+1 }{1-\gamma }}$ when $\gamma < -1 $ and tends to a constant if $\gamma=-1$. The fact that $M_0 $ is non-decreasing is not compatible with equation \eqref{eq:coag with growth}, where  only coagulation and growth take place, and hence we would expect the zero-th moment to decrease in time.

 \subsection{Overview of the results obtained in this paper}\label{sec:overview}
The goal of this paper is to analyse the long-time behaviour of the solutions to equation \eqref{S1E1} under the assumption 
%that the kernel $K$ is symmetric, homogeneous and satisfies \eqref{S1E7} with $F$ as in \eqref{S1E8}.  We also recall that it  satisfy the non-gelling conditions \eqref{non-gelling assumption} and
that $\gamma $ and $\lambda$ satisfy $\gamma+2\lambda \geq 1 $ and $\gamma \leq -1 $, which complements the results discussed above in Section \ref{sec:ExisResults} for non-gelling kernels. %\ref{sec:standardselfsim} and \ref{sec:ballisticselfsim}.

\subsubsection*{Anomalous self-similarity in the case $\gamma+2\lambda >1 $ and $\gamma \leq -1 $}
\begin{figure}[h]%
\centering
\includegraphics[width=1.2\textwidth]{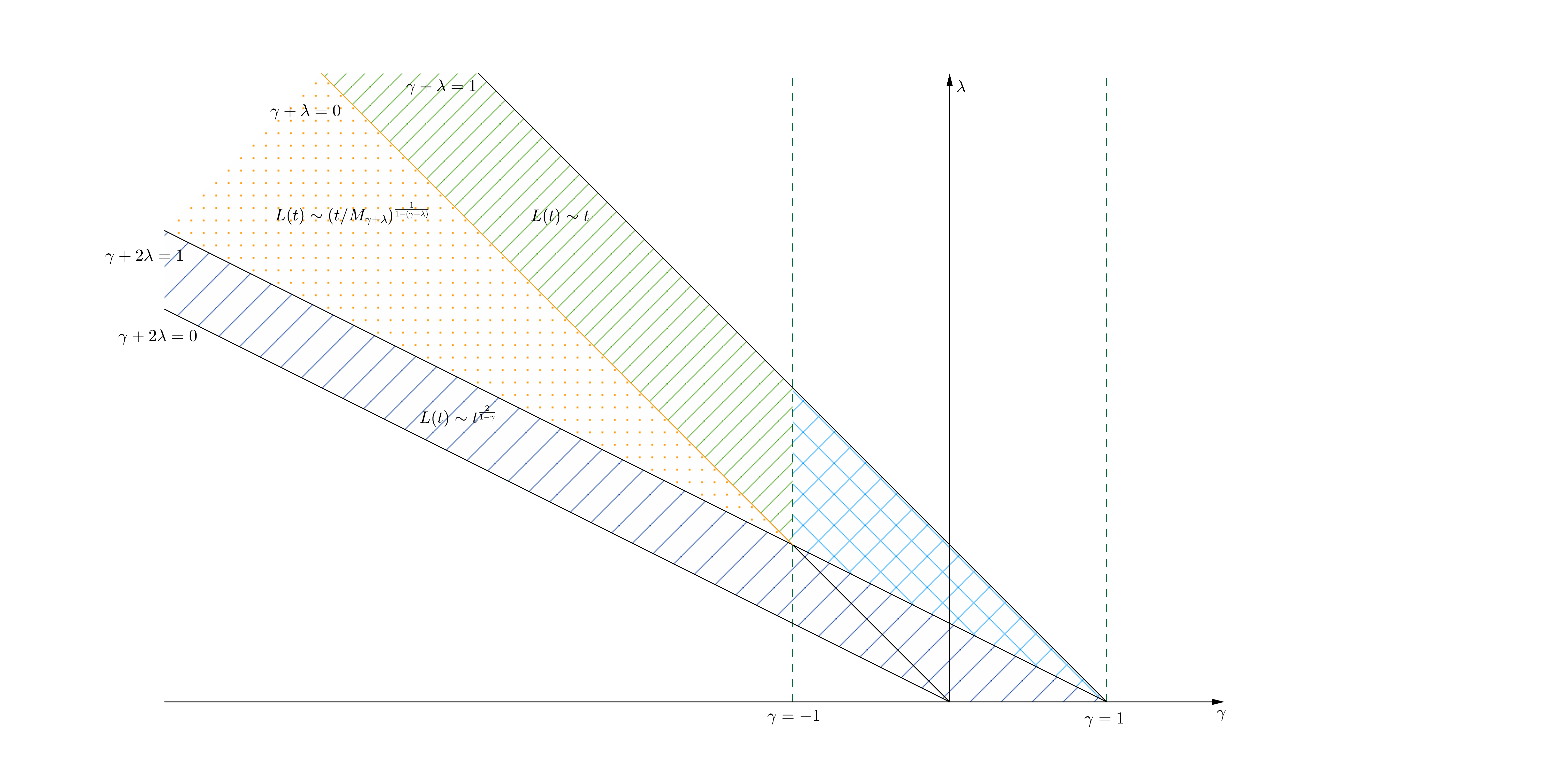}
\caption{Dependence on time of the characteristic length as the parameters $\gamma$ and $\lambda$ change. }
\label{fig1}
\end{figure}

As we will see in Section \ref{sec:dimensional}, when $\gamma \leq -1$ and $\gamma+2\lambda>1$ equation \eqref{eq:coag with growth} has some \textit{anomalous} scaling properties that depend also on the sign of $\gamma+\lambda$.
See Figure \ref{fig1} for a visual representation of how the scale $L(t)$ changes depending on the values of $\gamma$ and $\lambda$. 

We then expect that anomalous self-similar solutions exist for this equation; these solutions will be studied in Section \ref{sec:gamma+lambda>0} (when  $\gamma+\lambda >0$ and $\gamma < -1 $),  in Section \ref{sec:gaamma+lambda<0} (when  $\gamma+\lambda \leq 0$ and $\gamma < -1 $), and in Section \ref{sec:gamma=-1} (when  $\gamma=-1$).  
The precise notion of how the long-time behaviour of the solutions $f$ of equation \eqref{S1E1} become anomalous will be described at the beginning of each subsection.

%Here we use the term anomalous self-similar solutions to indicate that are self-similar solutions that do not correspond to the standard scaling properties that hold when $\gamma+2\lambda <1 $ and $\gamma >-1$, namely they do not have the form \eqref{standard self similar solution}.

In all the cases considered in Sections \ref{sec:gamma+lambda>0}, \ref{sec:gaamma+lambda<0} and \ref{sec:gamma=-1},  dimensional analysis considerations, that will be presented in detail in Section \ref{sec:dimensional}, imply that the moment $M_{\gamma+\lambda}$ tends to infinity as time tends to infinity.  
However, as will be explained in Section \ref{sec:regions}, we expect that the solution $f$ to equation \eqref{S1E1} approaches, in the inner region (cf.\ Section \ref{sec:regions}) of sizes of order $1$, a so-called {\it quasi-stationary solution} to equation \eqref{S1E1}, i.e., a solution to
\begin{equation} \label{B6}
\mathbb{K}\left[  f\right]  \left(  x\right)  -M_{\gamma+\lambda
}f\left(  x\right)  x^{-\lambda}+\eta\left(  x\right)  =0.
\end{equation}
We say that the solutions to this equation are  quasi-stationary because $M_{\gamma+\lambda}=M_{\gamma+\lambda}(f)$ depends on time, namely, it increases in time.  

When $M_{\gamma+\lambda } \rightarrow \infty $ for large times and for $\gamma+2\lambda >1 $ we obtain in Section \ref{sec:dimensional source} that the inner region produces a flux of clusters to the outer region (cf.\ Section \ref{sec:regions}).
This flux can be computed (cf.\ Section \ref{sec:inner region gamma+lambda to infinity}) analysing equation \eqref{B6} and depends on the $M_{\gamma+\lambda}$ moment in the following way 
\begin{equation}
J_c\left(  M_{\gamma+\lambda}\right)  \simeq 
KM_{\gamma+\lambda}\exp\left(  -a\left(  M_{\gamma+\lambda}\right)  ^{\frac
{2}{\gamma+2\lambda}}\right)  \label{MonFlux}%
\end{equation}
where $a$ and $K$ are suitable positive constants. 
 A similar analysis has been performed in  \cite{KMR98} and  \cite{KMR99} in the regime $\gamma\le -1$, $\gamma+\lambda=0$  where  asymptotic formulas which are consistent with \eqref{MonFlux} have been obtained. 

When $\gamma+2\lambda >1 $, $\gamma+\lambda >0$ and when $\gamma+2\lambda >1 $, $\gamma=-1$, we obtain that the flux of particles $J_c(M_{\gamma+\lambda})$ is negligible because $M_{\gamma+\lambda}$ tends to infinity polynomially. As we will see in Sections \ref{sec:gamma+lambda>0} and \ref{sec:gamma=-1}, this will lead to anomalous self-similarity with degenerated profiles.  
Instead, when $\gamma + 2 \lambda > 1 $ and $\gamma+\lambda \leq 0$ this flux of particles tends to infinity as the moment $M_{\gamma+\lambda}$ tends to infinity logarithmically. The profile in this case is obtained in Section \ref{sec:ss profile gamma+lambda<0}. 
According to the behaviour of $J_c(M_{\gamma+\lambda})$ and $M_{\gamma+\lambda}$ for different regimes we obtain different characteristic lengths  as shown in  Fig. \ref{fig1}.

\subsubsection*{Comments on the critical case $\gamma+2\lambda =1 $ % \mf{remove:} and $\gamma \leq -1 $
}
In this paper, we also consider the limiting case $\gamma+2\lambda=1$.  It 
appears to lead to interesting new possible asymptotic behaviour 
but it is also the case for which we have the least complete results.
Here, we will give a few preliminary results which allow us to formulate  
conjectures for possible behaviour in the limiting case.

It has been observed in  \cite{cristian2022long} that when $\gamma >-1$ and $\gamma+2\lambda=1$, a dimensional analysis argument implies that $M_{\gamma+\lambda} $ tends to a constant as time tends to infinity. 
Therefore, in this case, we do not expect the loss term of the coagulation operator to tend to infinity as time tends to infinity. Hence, we do not expect the solution $f$ to equation \eqref{S1E1} to tend to zero for small sizes. 
Instead, we expect that in the region of small sizes the solutions to equation \eqref{S1E1} behave like the solution to  equation \eqref{B6}  with a constant $M_{\gamma+\lambda}$,
where $M_{\gamma+\lambda} $ includes also the contribution from $f$ at large cluster sizes.   We refer to \cite{cristian2022long} and to Section \ref{sec:regions} for the details of the derivation of this equation. 
In particular, we recall that for these parameter values 
a self-similar solution with the standard scaling does exist, in the sense of Sec.\ \ref{sec:ballisticselfsim}.  This self-similar solution is expected to describe the long-time behaviour, in the region of large clusters, of the solutions $f$ to equation \eqref{S1E1}.

Interestingly, the existence/non-existence of solutions to equation \eqref{B6} is an open problem. In this paper, we present some formal arguments that suggest that there exists a critical value $M_{\text{crit}}$ such that when $M_{\gamma+\lambda} < M_{\text{crit}} $ a solution does not exist, while when $M_{\gamma+\lambda} \geq M_{\text{crit}}$, a solution exists. Note that this is compatible with the results in \cite{ferreira2019stationary}, where it is proven that when $\gamma + 2\lambda =1 $ a solution to \eqref{B6}
 does not exist if $M_{\gamma+\lambda} =0$.

% Another results that is proven in \cite{cristian2022long} is  that even when $\gamma + 2 \lambda =1 $ and $\gamma > -1 $ a standard self-similar solution exists. 
% This self-similar solution is expected to describe the long-time behaviour, in the region of large clusters, of the solutions $f$ of equation \eqref{S1E1}. 

For the complementary case, $\gamma \leq -1 $ and $\gamma+2\lambda=1$, we also conjecture that the solutions $f$ are well approximated by the solutions of \eqref{B6} at small cluster sizes. On the other hand, for large sizes,
 it has been proven in \cite{cristian2022long} that there are no
 self-similar solutions with the additional property
  \[
    \int_0^1 x^{-\lambda} \Phi(dx) < \infty \,.
    \]
This result is not conclusive: there could still be standard self-similar solutions such that $\int_0^1 x^{-\lambda} \Phi(dx) = \infty$. In this paper, we conjecture that,  when $\gamma+2\lambda =1$ and $\gamma \leq -1$, standard self-similar solutions,  for equation \eqref{eq:coag with growth}, with $\int_0^1 x^{-\lambda} \Phi(dx) = \infty $ could exist. 
In Figure \ref{fig6} (cf.\ Section \ref{sec:gamma=-1 and gamma+2lambda=1}), we emphasize that the form of the self-similar solutions strongly depends on the sign of the parameter $1-(\gamma+2\lambda)$, and it should not come as a surprise that that the analysis of the critical case $\gamma+2\lambda =1 $ is highly non trivial.

\subsection{Splitting the size space into inner, matching and outer regions} \label{sec:regions} 
In this section we derive heuristically equations \eqref{eq:coag with growth}  and \eqref{B6} describing, respectively, the behaviour of $f$ in the outer and 
the inner region. 
Here, we only sketch the results, as similar computations are presented in detail in \cite{cristian2022long}. 

\begin{figure}[h]%
\centering
\includegraphics[width=0.6\textwidth]{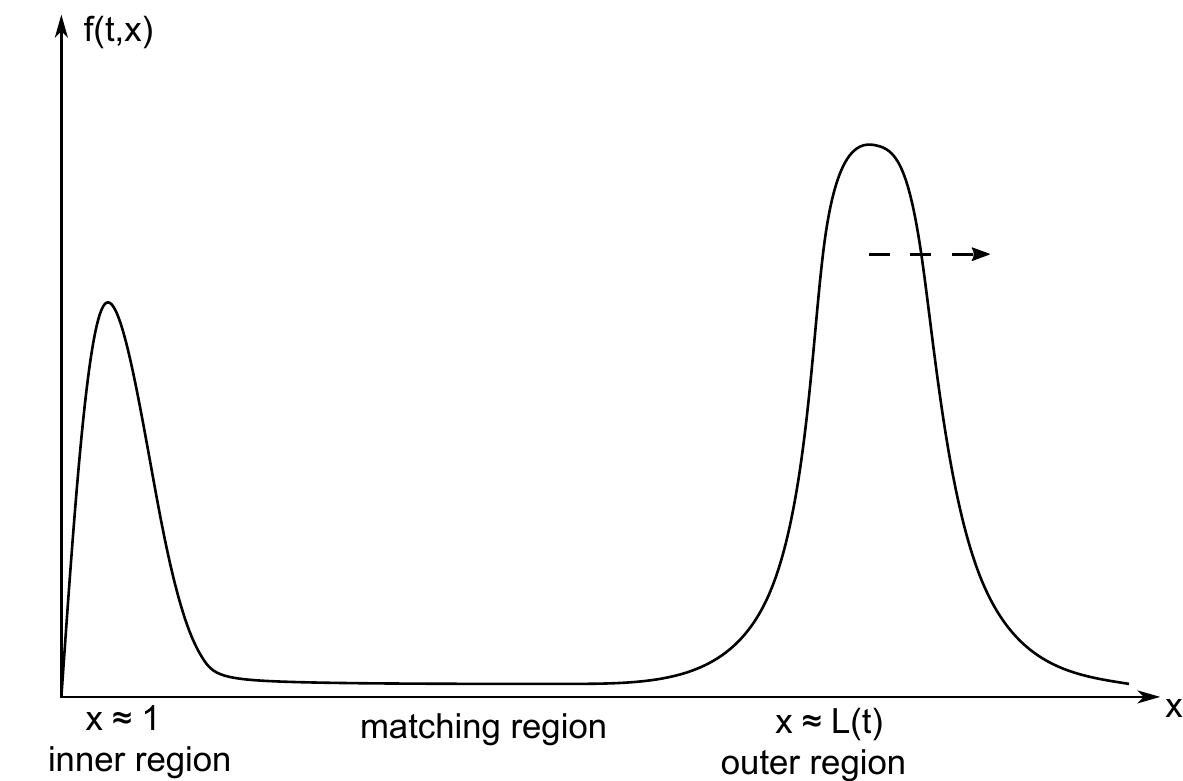}
\caption{The inner, the matching and the outer region are illustrated.}
\label{fig2}
\end{figure}

As a first step, we rewrite equation \eqref{S1E1} as a continuity equation
for the ``mass'' density $x f(t,x)$.  Namely,
if $f$ is a solution to \eqref{S1E1}, then
\begin{equation}
\partial_{t}\left(  x f(t,x)\right)+ \partial_{x}J[f(t,\cdot)]  (x) =x\eta\left(  x\right)  \label{fluxStat}%
\end{equation}
where the mass current observable, $J$, can be defined by
\begin{equation}\label{eq:flux}
J[f(t, \cdot) ](x) :=\int_{0}^{x}dy\int_{x-y}^{\infty}dz K(y,z)
y f\left( t,  y\right)  f\left( t,  z\right).
\end{equation}
Integrating over the $x$ variable, we get the corresponding continuity equation with source term. Assuming that the total injection rate has been scaled to one, we find
\[
\partial_t \int_0^z x f(t,x) dx + J[f(t, \cdot) ](z)=1, \qquad z > L_\eta\,.
\]
We then split the solution $f$ into three parts, $f= f_{in}+f_{m} + f_{out} $, with $f_{in} =f_{(0, \ell) },$
$f_{m} =f_{(\ell, L_0 ) }$ and $f_{out} = f_{(L_0, \infty ) }$.
Here, $L_0:= \epsilon L(t)$ is some suitably chosen characteristic length 
which guarantees that most of the mass lies in $f_{out}$ (see Section \ref{sec:dimensional}), while $1 \ll \ell \ll L_0(t) $. 
See Figure \ref{fig2} for an illustration of the inner, matching and outer regions. 

%Inserting this split
Using this decomposition in the definition of the flux $J$, \eqref{eq:flux}, yields
\[
 J[f(t, \cdot) ](z) = \int_0^z \int_{ z-x}^\infty xK(x,y) \left[ f_{in}(x)+f_m(x)+f_{out}(x)\right] \left[ f_{in}(y) +f_m(y)+f_{out}(y) \right]dy dx \,.
\]
We can then analyse each of the resulting terms using the asymptotic behaviour of the kernel $K$, as well as using the fact that we expect a stationary solution for $x \ll L_0(t) $.  This results in  equations for $f_{in}$, $f_m$ and $f_{out}$. 

In particular, we deduce that $f_{in} $ satisfies 
\begin{equation} \label{eq:flux inner}
1= J[f_{in}](z) + M_{\gamma+\lambda} \int_0^z x^{1-\lambda } f_{in}(x) dx . 
\end{equation}
Equation \eqref{eq:flux inner} is just equation \eqref{B6} written in flux form, and we will thus assume that $\overline f =f_{in}$ is a solution to that equation.
Since we assumed that most of the mass lies in the outer region, we can here use 
$M_{\gamma+\lambda }(f) = M_{\gamma+\lambda}(f_{out})$ and we will assume  that $M_{\gamma+\lambda }(f_m) \ll M_{\gamma +\lambda }(f)$.  
A consequence of equation \eqref{eq:flux inner} is that we expect to have that 
\begin{equation} \label{relation in and out moments} 
M_{\gamma+\lambda} M_{1-\lambda}^{in} =1,
\end{equation} 
where we are using the notation $M_{1-\lambda}(f_{in}) =M_{1-\lambda}^{in}$ and \eqref{C6}.
We stress here that equation \eqref{eq:flux inner} is expected to describe the behaviour of $f$ in the inner region both when $\gamma+2\lambda >1 $ (hence when $M_{\gamma+\lambda} \rightarrow \infty $ as time tends to infinity) and when $\gamma+2\lambda=1 $ (hence when $M_{\gamma+\lambda}$ is constant in time). The behaviours  of the solutions to \eqref{eq:flux inner} in both cases are studied in Section \ref{sec:inner region}. 
The relation \eqref{relation in and out moments} is derived by considering $z $ going to infinity in equation \eqref{eq:flux inner}.

Instead, when $z \approx \ell  $, we deduce that 
\[
1= J[f_m](z) + M_{1-\lambda}^{in} z^{\gamma+\lambda+1 } f_m (z) +  M_{\gamma+\lambda}   \int_0^z x^{1-\lambda } f_{m}(x) dx.  
\]
This equation contains the coagulation term, due to the interaction of particles of comparable sizes of order $\ell $, the loss term, due to the interaction between particles of order $\ell $ and particles of order $L$ and it contains a growth term, due to the interaction of particles of order $\ell $ with particles of order $1$. 
Finally, we remark that the term $1$ in the right-hand side of the above equation represents the flux of mass due to the source. 

Finally, we have that $f_{out} $ satisfies 
\[
\partial_t \int_0^z x f_{out}(t,x) dx + M_{1-\lambda}^{in} z^{\gamma+\lambda+1} f(z)+ J[f_{out}(t, \cdot)] (z)=1 \quad \text{ where } z \approx L(t).
\] 
Notice that \eqref{relation in and out moments} implies that this equation is just the flux formulation of equation \eqref{eq:coag with growth}, as expected.

\section{Characteristic length and main hypotheses when $\gamma+2\lambda >1 $} \label{sec:dimensional}
In this Section, we determine the different possible qualitative behaviours for the
solutions to (\ref{S1E1}) combining dimensional analysis arguments with some
estimates for the evolution of the moments $M_{0}$ and $M_{1}$ which yield the
behaviour of the total number of clusters and the total mass in
the system, respectively.
We start the section by recalling the dimensional analysis for the coagulation equation without source, then in Section \ref{sec:dimensional source} we will explain how these arguments change when we have the source. 

The main ansatz that we make in this paper 
%concerning the scaling propertiesof the solutions
is to assume that there exists a characteristic length $L$
depending on $t$ which characterizes the typical size of the clusters for long-times.
Typicality here is with respect to the probability distribution $f(t,x)/M_0(t)$.
Since the expected value for the size of the cluster is then equal to $M_1/M_0$,
our goal is to choose $L=L(t)$ so that 
the following relation between $[M_1]$ and $[M_0]$ is satisfied
\begin{equation}
\left[  M_{1}\right]  =\left[  M_{0}\right]  \left[  L\right]\,,
\label{MomDimAn}%.
\end{equation}
up to some factor of order one on the right hand side.   After fixing the scale $L(t)$, we also assume that it then determines the scale of all other relevant moments $q$,
\[
\left[  M_{q}\right]  \approx \left[  M_{0}\right]  \left[  L\right]^q.
\]
This would certainly be true if the dominant contribution to the moment comes from a self-similar solution as in (\ref{standard self similar solution}).
Notice that \eqref{MomDimAn} implicitly assumes that $M_0(f_{out})\approx M_0(f)$ and that $M_0(f_{in})$ is negligible. Later we will see that these assumptions are self-consistent.

\subsection{Dimensional analysis without source}
\label{sec:dimensional without source}

In this first case, we assume  that there is no source and that gelation does not take place.  Hence, $[M_1] \approx 1 $. Together with \eqref{MomDimAn} this implies that $[M_0] = [L]^{-1} $. 
On the other hand, since $f$ satisfies the coagulation equation and since the kernel is homogeneous, we have the following ODE describing the evolution in time of the $0$-th moment
\begin{equation}\label{evolution number no source} 
    \frac{d}{dt} M_0 = -[M_0]^{2} [L]^{\gamma}. 
\end{equation}
Using the relation between $[M_0]$ and $L$, equation \eqref{evolution number no source} can be rewritten as $    \frac{d}{dt} [M_0] = -[M_0]^{2-\gamma }.$
Note that since $\gamma<1 $, the number of particles is always decreasing in time, as expected. As we will see later this will, instead, depend on the assumptions on the parameters $\gamma $ and $\lambda $ when we consider the coagulation equation with source. 
Solving the ODE for $[M_0]$, we deduce that $[M_0] \approx t^{-\frac{1}{1-\gamma}}$. Hence the characteristic length is $L \approx t^{\frac{1}{1-\gamma} }$.
We will next see how this argument changes if we consider the coagulation equation with a source. 

\subsection{Dimensional analysis with  source} \label{sec:dimensional source}

In this section, we will consider only the case $\gamma+2\lambda >1 $ as the case $\gamma+2\lambda =1 $ is different and more challenging, therefore it will be considered separately in Section \ref{sec:gamma=-1 and gamma+2lambda=1}. 

Since we are in the non-gelling regime and we are considering the coagulation equation with source  we deduce that the  first
moment evolves according to the following ODE
\[
\frac{dM_{1}}{dt}=\int_{\mathbb R_*} x \eta(x) d x=1\,. 
\]
Here, we have adapted the microscopic time-scale so that the total injection rate from the source leads to rate one injection of the total mass.   This scaling assumption will be continued to be used from now on.

In the last equality comes from our assumption that 
Therefore,%
\[
M_{1}(t)=M_{1,0}+t
\]
where $M_{1,0}$ is the initial mass of monomers.
We then have the following asymptotic behaviour for long-times%
\begin{equation}
M_{1}\approx t\ \ \text{as\ \ }t\rightarrow\infty.\label{M1resc}%
\end{equation}
As in Section \ref{sec:dimensional without source}, the dimensional analysis, combined with our assumption on the
existence of a characteristic length $L$, 
will allow to assume \eqref{MomDimAn}. Thanks to (\ref{M1resc}),
this results in the following relation
between $\left[  M_{0}\right]  $ and the characteristic length $L$%
\begin{equation}
\left[  M_{0}\right]  \left[  L\right]  =t. \label{Mom0Lrel}%
\end{equation}
Notice that to deduce equations \eqref{M1resc} and \eqref{Mom0Lrel}, we only used the non-gelling conditions for $\gamma$ and $\lambda$, hence these asymptotics hold for any value of $\gamma+\lambda < 1$ and $\gamma < 1$.  

We now compute the derivative of $M_{0}.$ We use the assumption that most of the
clusters are in the region where $x\gg1$  and therefore that $M_{0}$ is given
approximately by the integral (or sum) on this region of cluster sizes, i.e., $M_0\approx M_0(f_{out})$.

Using the homogeneity of the kernel, we can write
the dimensions of the coagulation operator $\mathbb{K}\left[  f\right]  $ in
(\ref{S1E1}) as $\left[  M_{0}\right]  ^{2}\left[  L\right]  ^{\gamma-1}.$
Moreover, each coagulation event reduces the number of clusters by one. On the
other hand, there is an implicit source of clusters %due to the presence of the term $\eta\left(  x\right)  $ 
in equation \eqref{eq:coag with growth}, represented by the growth term. 
The added clusters can be transferred to large
cluster sizes by means of  different mechanisms 
%depending on the sizes of the coagulating particles,
which may or may not  contribute to an increase of $M_0$. 

One mechanism consists in
the coagulation of a cluster with size of order one with a large cluster.
However, such a transfer mechanism does not modify the number of large
clusters $M_{0}.$
On the contrary the coagulation of two large clusters
results in the reduction of one cluster by reaction. Therefore, the
contribution of the coagulation mechanism to $\frac{dM_{0}}{dt}$ is $-\left[
M_{0}\right]  ^{2}\left[  L\right]  ^{\gamma},$ where we use the fact that
this term yields the order of magnitude of $\int_{0}^{\infty}\mathbb{K}\left[
f\right]  dx.$ 
Finally, the number of large clusters $M_{0}$ can increase due
to the flux of small clusters to large clusters, due to iterated coagulation
events of the small injected clusters. The flux of clusters from small cluster
to large cluster sizes has been computed in (\ref{MonFlux}).

We then have the
following equation that yields the order of magnitude for the change of
$M_{0}$%
\[
\frac{dM_{0}}{dt}=J_c\left(  M_{\gamma+\lambda}\right)  -\left[  M_{0}\right]
^{2}\left[  L\right]  ^{\gamma}, 
\]
We notice that this equation can be obtained from \eqref{eq:coag with growth} by integrating over the size variable and taking into account the influx of particles from zero, represented here by $J_c(M_{\gamma+\lambda})$.
Combining the previous formula with (\ref{Mom0Lrel}) we obtain the following equation
for $M_{0}$%
\begin{equation}
\frac{dM_{0}}{dt}=J_c\left(  M_{\gamma+\lambda}\right)  -\left[  M_{0}\right]
^{2-\gamma}t^{\gamma}.  \label{diffEquMon}%
\end{equation}

The equation (\ref{diffEquMon}) allows to obtain many different behaviours for
the solutions for different choices of the exponents $\gamma$ and $\lambda.$
Let us point out that the type of behaviours allowed by
(\ref{diffEquMon}) is much richer than the ones obtained for the coagulation
equation without injection \eqref{evolution number no source}. A major difference with the problem without
injection is the fact that in the problem with injection we can have both
increasing and decreasing number of clusters $M_{0}.$ In the case of
coagulation without injection the number of clusters $M_{0}$ can only decrease
due to the effect of the coagulation.

By  \eqref{MonFlux},
the fluxes $J_c\left(  M_{\gamma+\lambda}\right)  $ depend
exponentially in $M_{\gamma+\lambda}$. 
Our assumption about the characteristic
length $L$ indicates that $\left[  M_{\gamma+\lambda}\right]  =\left[
M_{0}\right]  \left[  L\right]  ^{\gamma+\lambda}.$ Now if
$M_{\gamma+\lambda}$ tends to infinity as a power law of $t$, the flux of
monomers $J_c\left(  M_{\gamma+\lambda}\right)  $ would be exponentially small
and therefore the gain of monomers would be negligible. In such a case,  the number of clusters evolves according to the equation%
\begin{equation}
\frac{dM_{0}}{dt}=-\left[  M_{0}\right]  ^{2-\gamma}t^{\gamma}\label{mom0decr}%
\end{equation}

\begin{figure}[h]%
%\centering
\hspace{-2cm}\includegraphics[width=1.3\textwidth]{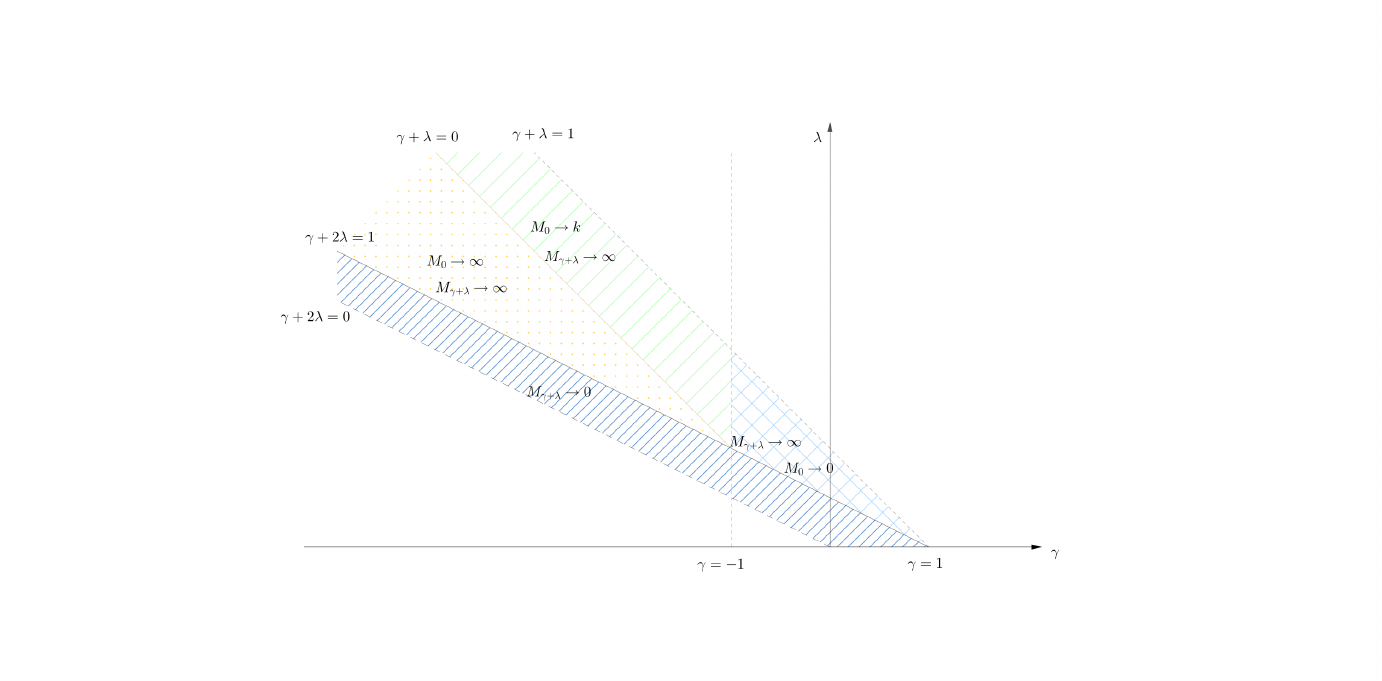}
\caption{In this picture we show how the behaviour of the moments $M_{\gamma+\lambda}$ and $M_0$ change as the parameters $\lambda$ and $\gamma$ change. }
\label{fig3}
\end{figure}

We stress that the approximation (\ref{mom0decr}) can be expected to be valid
only if $M_{\gamma+\lambda}$ tends to infinity sufficiently fast, for instance,
like a power law.  Failing that,  the analysis of \eqref{diffEquMon} suggests   different behaviours for $M_{0}$.
We refer to Figure \ref{fig3} for a summary of how $M_0$ and $M_{\gamma+\lambda}$ change depending on the parameters $\gamma $ and $\lambda$. 
The results from the analysis of the various scenarios in the regime $\gamma+2\lambda >1$ are summarized in the following:

\bigskip

\begin{itemize}
\item If $\gamma>-1$, we obtain from (\ref{mom0decr}) the asymptotics $\left[
M_{0}\right]  \approx t^{-\frac{1+\gamma}{1-\gamma}}.$
Then, the number of
clusters converges to zero and (\ref{Mom0Lrel}) implies $L\approx t^{\frac
{2}{1-\gamma}}.$ This corresponds to the self similar behaviour that has been
considered in \cite{cristian2022long}, see equality \eqref{standard self similar solution} for the self-similar solution. 
Note that in this case we
have $\left[  M_{\gamma+\lambda}\right]  =\left[  M_{0}\right]  \left[
L\right]  ^{\gamma+\lambda}\approx t^{\frac{\gamma+2\lambda-1}{1-\gamma}}$
that converges to infinity as a power law, since $\gamma+2\lambda>1$.  
This implies that  $J_c\left(  M_{\gamma+\lambda}\right)  $ converges to zero
exponentially fast, justifying the approximation (\ref{mom0decr}) and
therefore asserting the self-consistency of this approximated equation.

\item If $\gamma=-1$, we obtain the asymptotic behaviour $\left[  M_{0}\right]
\approx\left(  \log\left(  t\right)  \right)  ^{-\frac{1}{2}}.$ Using
(\ref{Mom0Lrel}) we obtain $L\approx t\left(  \log\left(  t\right)  \right)
^{\frac{1}{2}}$ as $t\rightarrow\infty.$ This corresponds to a class of
self-similar solutions with a logarithmic correction that will be described in
Section \ref{sec:gamma=-1}. Note that in this case we have \[
\left[  M_{\gamma+\lambda
}\right]  \approx\left(  \log\left(  t\right)  \right)  ^{\frac{1}{2}\left(
\gamma+\lambda-1\right)  }t^{\gamma+\lambda}=\left(  \log\left(  t\right)
\right)  ^{\frac{1}{2}\left(  \lambda-2\right)  }t^{\lambda-1} \text{ as } t\rightarrow\infty.
\]
Since $\gamma+2\lambda>1$ and $\gamma=-1$, we have here
$\lambda>1.$ Hence, $\left[  M_{\gamma+\lambda}\right]  $ tends to infinity like
a power law. Therefore, given \eqref{MonFlux}, $J_c\left(  M_{\gamma+\lambda}\right)  $ is exponentially
small, showing the consistency of the approximation (\ref{mom0decr}).
\item If  $\gamma<-1$ we have two possibilities: 
\begin{itemize}
\item If  $\left(  \gamma+\lambda\right)  >0$, \eqref{mom0decr} implies that
$M_{0}$ approaches to a constant as $t\rightarrow\infty.$ 
We expect the number of clusters to tend to a constant for large times, because the rate of
destruction of clusters due to the coagulation process is too slow to make the number of cluster decreasing and the flux of particles from the origin is too slow to make it increase.  
Using (\ref{Mom0Lrel}) it follows that $L\approx t.$ Then $\left[  M_{\gamma
+\lambda}\right]  \approx t^{\gamma+\lambda}.$ Therefore, if $\left(
\gamma+\lambda\right)  >0$, we obtain that the fluxes $J_c\left(  M_{\gamma
+\lambda}\right)  $ are exponentially small and the approximation
(\ref{mom0decr}) is self-consistent.

\item 
If $\left(  \gamma+\lambda\right)  \leq 0$, dimensional
considerations do not imply $\left[  M_{\gamma+\lambda}\right] $ approaching polynomially
to infinity for large times. Therefore, due to \eqref{MonFlux} we can expect to have a nontrivial
contribution to the number of clusters $M_{0}$ due to the fluxes of small
clusters to large clusters $J_c\left(  M_{\gamma+\lambda}\right)$. Due to the
exponential dependence of the fluxes $J_c\left(  M_{\gamma+\lambda}\right)  $ in
$M_{\gamma+\lambda}$, the only way in which we can avoid these fluxes to be
exponentially small is by assuming that they increase at most in a logarithmic
manner in $t$ (by construction with a power law behaviour) as $t\rightarrow
\infty.$ 
Using that $\left[  M_{\gamma+\lambda}\right]
=\left[  M_{0}\right]  \left[  L\right]  ^{\gamma+\lambda}$ and using also
(\ref{Mom0Lrel}) it follows that 
\begin{equation}\label{L char} 
    L^{1-\left(  \gamma+\lambda\right)} %
\approx \frac{  t }{[M_{\gamma+\lambda}]}.
\end{equation}
Then $[M_{0}] \cdot t^{\frac{\left(  \gamma+\lambda\right)
}{1-\left(  \gamma+\lambda\right)  }}\approx [M_{\gamma+\lambda}]^{\frac{1}{1-( \gamma+\lambda) }} .$
Therefore, the contribution of the coagulation term to $-\frac{dM_{0}}{dt}$ is of the order $\left[  M_{0}\right]  ^{2-\gamma}t^{\gamma}\approx \frac{\left[  M_{0}\right] [M_{\gamma+\lambda}]^{\frac{1-\gamma}{1-(\gamma+\lambda) }} }%
{t^{\frac{\lambda}{1-\left(  \gamma+\lambda\right)  }}}.$ 
Since $M_{\gamma+\lambda}$ grows logarithmically, we have that, when $\gamma+2\lambda>1$, it then follows that this term is much
smaller than $\frac{dM_{0}}{dt}.$ 
Therefore (\ref{diffEquMon}) reduces to%
\[
\frac{dM_{0}}{dt}=J_c\left(  M_{\gamma+\lambda}\right) .
\]
Using now (\ref{MonFlux}) we can obtain the logarithmic dependence of
$M_{\gamma+\lambda}.$ Indeed, using that $[M_{0}]\approx [M_{\gamma+\lambda}]^{\frac{1}{1-(\gamma+\lambda)}}\cdot
t^{-\frac{\left(  \gamma+\lambda\right)  }{1-\left(  \gamma+\lambda\right)  }%
}$ it then follows that%
\begin{equation} \label{asymp of Mgamma+lambda}
M_{\gamma+\lambda}\simeq\left(  \frac{1}{a_{\gamma+2\lambda}}\log\left(  t^{\frac{1}{1-\left(
\gamma+\lambda\right)  }}\right)  \right)  ^{\frac{\gamma+2\lambda}{2}%
}\ \ \text{as\ \ }t\rightarrow\infty.
\end{equation} 
We stress that the constant $a_{\gamma+2\lambda}$ in \eqref{asymp of Mgamma+lambda} depends on the value of $\gamma+2\lambda$. 

Combining this formula with $\left[  M_{\gamma+\lambda}\right]  =\left[
M_{0}\right]  \left[  L\right]  ^{\gamma+\lambda}$ and (\ref{Mom0Lrel}) we can
obtain the scaling laws for $M_{0}$ and $L$ (including logarithmic terms).
These scalings will allow to obtain approximate solutions to the coagulation
equation, that will be determined in Section \ref{sec:gaamma+lambda<0}.

\end{itemize}
\end{itemize}

The dimensional considerations above include all the cases for which standard self-similar solutions to equation \eqref{S1E1} do not exist for homogeneous kernels of the form \eqref{S1E7} with $F$ satisfying \eqref{S1E8} and the non-gelation conditions \eqref{non-gelling assumption}.
%In the following sections, we present the formal asymptotic analysis of solutions to equation \eqref{S1E1} that  augment 
 %\textcolor{red}{A: I would suggest to slightly rephrase the previous sentence, for instance: ``In the following sections we provide the formal computations, based on asymptotic analysis techniques, that allow to justify the asymptotics of solutions"} presented in \cite{cristian2022long}, in \cite{ferreira2021self} and in this paper. 
In the following sections we provide the formal computations, based on asymptotic analysis techniques, that allow to justify the long-time behaviour of the solutions to \eqref{S1E1}.
 
%\newpage

\section{Behaviour in the inner region} \label{sec:inner region} 
In this section we analyse the long-time behaviour of the solutions $f$ to equation \eqref{S1E1} in the region of sizes $x$ of order $1$. 
We assume that the kernel $K$ satisfies \eqref{S1E7} and \eqref{S1E8} with parameters that satisfy the non-gelation condition \eqref{non-gelling assumption}  and that satisfy $\gamma + 2 \lambda \geq 1 $. 

As already anticipated in the introduction, we expect two different behaviours depending on the long-time behaviour of the $M_{\gamma+\lambda} $ moment. 
As we discussed in Section \ref{sec:dimensional}, dimensional considerations imply that when $\gamma+2\lambda > 1 $, then $M_{\gamma+\lambda}$ tends to infinity as time tends to infinity (although in different ways depending on the signs of $\gamma + 1$ and of $\gamma+\lambda $). 
The long-time behaviour of the solution $f$ to equation \eqref{S1E1} in the region of small sizes corresponding to $M_{\gamma+\lambda} \rightarrow \infty $ as $t \rightarrow \infty $ is studied in Section \ref{sec:inner region gamma+lambda to infinity}.  

When, instead, $\gamma+2\lambda =1 $  and $\gamma >-1$, dimensional considerations show that $M_{\gamma+\lambda} $ tends to a constant as time goes to infinity. For more details, we refer to \cite{cristian2022long} and to Section \ref{sec:dimensional}. 
When $\gamma+2\lambda=1 $ and $\gamma \leq -1 $, we also expect that $M_{\gamma+\lambda}$ tends to a constant and therefore we expect the same behaviour as in the case $\gamma+2\lambda =1 $ and $\gamma > -1 $ in the inner region. 
%For this case we will present  only preliminary results as a rigorous analysis of this case is challenging and goes beyond the aim of this paper. 
For this set of parameters, the long-time behaviour of the solution $f$ to equation \eqref{S1E1} in the region of small sizes is studied in Section \ref{sec: inner behaviour when M constant}.

\subsection{Behaviour of the solutions to \eqref{B6} %\eqref{eq:coag with growth} 
when $M_{\gamma+\lambda} \rightarrow \infty $ as $t \rightarrow \infty$} \label{sec:inner region gamma+lambda to infinity} 
\bigskip

The goal of this Section is to obtain the behaviour of the solutions to \eqref{B6} when the moment $M_{\gamma+\lambda}\to \infty$ for large times. In order to do this, we  first recall that there exists a characteristic cluster size $L\left(  t\right)  $
satisfying $\lim_{t\rightarrow\infty}L\left(  t\right)  =\infty$ such that
most of the monomers of the system are contained in the clusters with sizes of
order $ L\left(  t\right) .$

We make some  additional assumption which will allow to simplify the problem. 
\begin{itemize}
\item[(i)]  The cluster distribution $f(t,x)$ can be approximated by means of a discrete
distribution $c_n(t)$ for small values   of $x.$ 
\item[(ii)] We will assume that $\eta
_{n}=a\delta_{n,1}$ for some $a>0.$ 
Rescaling the values of $c_{n}$ and
changing the unit of time we can assume without loss of generality that
$a=1.$
\end{itemize}

%Moreover, the concentrations of clusters $c_n(t)$ of order one can be approximated by a quasistationary cluster distribution.
%More precisely, we can approximate the concentrations $c_{n}\left(  t\right)$ for $n\approx1$ using the equation%
%\begin{equation}
%\frac{1}{2}\sum_{j=1}^{n-1}K\left(  n-j,j\right)  c_{n-j}c_{j}-\sum
%_{j=1}^{\infty}K\left(  j,n\right)  c_{j}c_{n}+\delta_{n,1}=0. \label{A7}%
%\end{equation}

%Notice that the distribution $f\left( t, x\right)$ must be thought as the ``outer"
%distribution of clusters and it should be matched with the
%``inner" concentration of clusters that is described by means of the equation
%(\ref{A7}). We remark that some of the cluster concentrations $c_{n}$
%appearing in (\ref{A7}) are described by means of the outer concentration
%$f\left( t, x\right).$ 
%In order to distinguish between the ``inner" clusters
%and the ``outer" clusters, we introduce a cutoff parameter %$L\left(  t\right)
%$ such that $\lim_{t\rightarrow\infty}L\left(  t\right)  %=\infty.$ We will
%denote as ``inner" clusters those with $1\leq n\leq L\left(  %t\right)  $ and
%as ``outer" clusters those with $n>L\left(  t\right).$ C
Combining the assumption on the existence of a characteristic size with the approximation $K\left(  j,n\right)  \simeq\left(  j\right)^{\gamma+\lambda
}\left(  n\right)  ^{-\lambda}$ for $n$ of order one and $j$ large, as well as, 
assumptions (i) and (ii) we obtain the following discrete version  of equation \eqref{B6}  for the
concentrations $c_{n}\left(  t\right)$ in the inner region, namely when $n\approx1$,  %( n of order one)
\begin{equation}
\frac{1}{2}\sum_{j=1}^{n-1}K\left(  n-j,j\right)  c_{n-j}c_{j}-\left[
\sum_{j=1}^{L\left(  t\right)  }K\left(  j,n\right)  c_{j}\right]  c_{n}%
-\frac{M_{\gamma+\lambda}}{\left(  n\right)  ^{\lambda}}c_{n}+\delta
_{n,1}=0.\label{A8a}%
\end{equation}
% Here we have also used %the fact that \eqref{A1}, \eqref{A2} imply  the approximation $K\left(  j,n\right)  \simeq\left(  j\right)^{\gamma+\lambda}\left(  n\right)  ^{-\lambda}$ for $n$ of order one and $j$ large.
We mention that a similar strategy has been used for a   Becker--D\"oring  model in \cite{niethammer2022oscillations}.

%We will assume also that $f$ and $L\left(  t\right)  $ are such that $\int_{L\left(  t\right)  }^{\infty}\left(  x\right)  ^{\gamma+\lambda}f\left(  x,t\right)  dx$ can be approximated by the integral in the whole space, i.e., 
%$$
%M_{\gamma+\lambda}=\int_{0}^{\infty}\left(  x\right)  ^{\gamma+\lambda
%}f\left(  x,t\right)  dx.
%$$
%Notice that this approximation assumes implicitly that it is possible to define the outer cluster concentration $f\left( t, x \right)  $ up
%to small values of the continuous variable $x.$ 
%We recall that, in addition to the assumptions (i)-(iii) above, in this section we are assuming that the moment $M_{\gamma+\lambda}$ defined by means of (\ref{C6})
%satisfies %$\lim_{t\rightarrow\infty}M_{\gamma+\lambda}=\infty.$ 
%We remark that this convergence  can be very slow. As a matter of fact, in some of the cases considered later we will have that $M_{\gamma+\lambda}$ diverges logarithmically in $t.$ 

%Thanks to the assumptions above on $f$ and $L\left(  t\right)$ we can 
%approximate (\ref{A8}) by
%\begin{equation}
%\frac{1}{2}\sum_{j=1}^{n-1}K\left(  n-j,j\right)  c_{n-j}c_{j}-\left[
%\sum_{j=1}^{L\left(  t\right)  }K\left(  j,n\right)  c_{j}\right]  c_{n}%
%-\frac{M_{\gamma+\lambda}}{\left(  n\right)  ^{\lambda}}c_{n}+\delta
%_{n,1}=0\label{A8a}%
%\end{equation}
%which provides an equation for the concentrations $c_{n}$ for 
%the ``inner" clusters, namely for $n \approx 1 .$  
We make also the following additional assumption which allows to simplify (\ref{A8a}). 
\begin{itemize}
\item[(iii)] We consider only solutions to (\ref{A8a})  that satisfy
\begin{equation}
\sum_{j=2}^{L\left(  t\right)  }K\left(  j,n\right)  c_{j}\ll\max\left\{
K\left(  1,n\right)  c_{1},\frac{M_{\gamma+\lambda}}{\left(  n\right)
^{\lambda}}\right\}  \label{A8b}%
\end{equation}
as $M_{\gamma+\lambda}\rightarrow\infty$ uniformly for all the values of
$n\in\left[  1,L\left(  t\right)  \right].$ 
\end{itemize}
We notice that condition (iii) is motivated by the strong interaction between particles of size $1$ and $n$ that, as explained in the introduction, we expect to have when $\gamma+2\lambda \geq 1$. 
%Later we can check that this assumption is self-consistent with the asymptotics that we deduce. 
Assumption (iii) implies that we can approximate (\ref{A8a}) by%
\begin{equation}
\frac{1}{2}\sum_{j=1}^{n-1}K\left(  n-j,j\right)  c_{n-j}c_{j}-K\left(
1,n\right)  c_{1}c_{n}-\frac{M_{\gamma+\lambda}}{\left(  n\right)  ^{\lambda}%
}c_{n}+\delta_{n,1}=0.\label{A8c}%
\end{equation}
Notice that the equation (\ref{A8c}) can be solved recursively for arbitrary values of $n.$ 
Indeed, we can determine $c_{1}$ as the unique positive
solution to the equation%
\begin{equation}
K\left(  1,1\right)  \left(  c_{1}\right)  ^{2}+M_{\gamma+\lambda}%
c_{1}=1\label{A8da}%
\end{equation}

Therefore, we can determine $c_{n}$ for $n\geq2$ by means of the iterative
formula%
\begin{equation}
c_{n}=\frac{1}{2\left(  K\left(  1,n\right)  c_{1}+\frac{M_{\gamma+\lambda}%
}{\left(  n\right)  ^{\lambda}}\right)  }\sum_{j=1}^{n-1}K\left(
n-j,j\right)  c_{n-j}c_{j}\ \ \ \ n\geq2.\label{A8d}%
\end{equation} 
It is then possible to prove that the solutions to (\ref{A8da}), (\ref{A8d}) satisfy the following asymptotics
\begin{equation}
c_{n}\simeq\frac{\left[  \left(  2\pi\right)  ^{\frac{\gamma}{2}+\lambda
}K\right]  \left(  M_{\gamma+\lambda}\right)  ^{2}}{n^{\gamma+\lambda}%
\sqrt{1+\frac{\left(  M_{\gamma+\lambda}\right)  ^{2}}{n^{\gamma+2\lambda}}}%
}\exp\left(  -\left(  M_{\gamma+\lambda}\right)  ^{\frac{2}{\gamma+2\lambda}%
}\int_{0}^{\frac{n}{\left(  M_{\gamma+\lambda}\right)  ^{\frac{2}%
{\gamma+2\lambda}}}}\log\left(  1+\frac{1}{y^{\gamma+2\lambda}}\right)
dy\right)  \label{C7}%
\end{equation}
as $M_{\gamma+\lambda}\rightarrow\infty$ and $1\ll n\ll L\left(  t\right)$ for a
suitable constant $K>0$ independent of $M_{\gamma+\lambda}.$

\subsubsection*{Strategy for the justification of \eqref{C7}}
In this section we only give the main ideas for the computation of \eqref{C7}, some more details can be found in the appendix \ref{appendix}. These ideas are very similar to the ones in \cite{KMR98}, \cite{KMR99}. 
First of all notice that, if $M_{\gamma+\lambda}\rightarrow\infty$, we can approximate the first values of the sequence $\left\{
c_{n}\right\}  _{n\in\mathbb{N}}$ solving (\ref{A8da}), (\ref{A8d}). We
have%
\begin{equation}
c_{1}\simeq\frac{1}{M_{\gamma+\lambda}}\ ,\ \ c_{2}\simeq\frac{1}{2}%
\frac{2^{\lambda}K\left(  1,1\right)  }{\left(  M_{\gamma+\lambda}\right)
^{3}}\ ,\ \ c_{3}\simeq\frac{1}{2}\frac{3^{\lambda}2^{\lambda}K\left(
2,1\right)  \left(  K\left(  1,1\right)  \right)  ^{2}}{\left(  M_{\gamma
+\lambda}\right)  ^{5}}\ ,\ ... \label{B5}%
\end{equation}
We now want to approximate the values of $\left\{  c_{n}\right\}  _{n\in
\mathbb{N}}$ using the approximation given by (\ref{A8d}) for a range of values
of $n$ for which the term $K\left(  1,n\right)  \left(  n\right)  ^{\lambda
}c_{1}$ becomes comparable or larger than $M_{\gamma+\lambda}.$  Notice that, due to the asymptotics of $c_{1}$ in
(\ref{B5}) we have that $K\left(  1,n\right)  \left(  n\right)  ^{\lambda
}c_{1}$ becomes comparable to $M_{\gamma+\lambda}$ if $K\left(  1,n\right)
\left(  n\right)  ^{\lambda}c_{1}\simeq\frac{\left(  n\right)  ^{\gamma
+2\lambda}}{M_{\gamma+\lambda}}\approx M_{\gamma+\lambda},$ i.e., for
\thinspace$n\approx\left(  M_{\gamma+\lambda}\right)  ^{\frac{2}%
{\gamma+2\lambda}}.$

We would like to check that the main contribution to the convolution term
$\sum_{j=1}^{n-1}K\left(  n-j,j\right)  c_{n-j}c_{j}$ is due to the terms
multiplying $c_{1}$ if $t\rightarrow\infty$ and $1\ll n\leq L\left(  t\right)
.$ We first check this for the range of values of $n$ for which $K\left(
1,n\right)  \left(  n\right)  ^{\lambda}c_{1}\ll M_{\gamma+\lambda},$ i.e.,
$n\ll\left(  M_{\gamma+\lambda}\right)  ^{\frac{2}{\gamma+2\lambda}}.$ In this
case, we can approximate (\ref{A8d}) as%
\begin{equation} \label{cn approx}
c_{n}=\frac{\left(  n\right)  ^{\lambda}}{M_{\gamma+\lambda}}\left[  \frac
{1}{2}\sum_{j=1}^{n-1}K\left(  n-j,j\right)  c_{n-j}c_{j}+\delta_{n,1}\right]
\ \ ,\ \ n\geq1. 
\end{equation}

The asymptotic formulas (\ref{B5}) suggest to rescale the concentrations
$c_{n}$ as%
\begin{equation} \label{c_n}
c_{n}=\frac{X_{n}}{\left(  M_{\gamma+\lambda}\right)  ^{2n-1}}\ \ ,\ \ n\in
\mathbb{N}.%
\end{equation}
Using \eqref{cn approx} then we obtain the following equation for the variables $X_{n}$%
\begin{equation}
X_{n}=\frac{\left(  n\right)  ^{\lambda}}{2}\sum_{j=1}^{n-1}K\left(
n-j,j\right)  X_{n-j}X_{j}\ \ ,\ \ n\geq2\ \ ,\ \ X_{1}=1. \label{B62}%
\end{equation}

Notice that (\ref{B62}) allows to obtain the sequence of coefficients $\left\{
X_{n}\right\}  _{n\in\mathbb{N}}$ in an iterative manner. We claim that under
the assumption $\gamma+2\lambda\geq 1$  
the asymptotic behaviour of
the solutions to (\ref{B62})\ can be approximated by the solutions to the
problem 
\begin{equation}
X_{n}=\left(  n\right)  ^{\lambda}K\left(  n-1,1\right)  X_{n-1}%
X_{1}\ \ \ \text{for }n\text{ large}. \label{C2}%
\end{equation}
Using \eqref{S1E7}, \eqref{S1E8}, as well as the fact that $X_{1}=1$, we obtain
the leading order approximation 
\[
X_{n}=\left(  n\right)  ^{\gamma+2\lambda}X_{n-1}\ \ \text{for }n\text{ large}%
\]
and, finally we deduce the approximation
\begin{equation}
X_{n}\simeq\frac{K}{n^{\gamma+\lambda}}\left(  n!\right)  ^{\gamma+2\lambda
}\ \ \text{as }n\rightarrow\infty\label{C1}%
\end{equation}
for some suitable constant $K$. 
%Then when $n \approx M_{\gamma+\lambda}^{\frac{2}{\gamma+2\lambda}}$, combining \eqref{C1} with \eqref{c_n} and using Stirling's approximation for the factorial we infer that
%\eu{\begin{align*}
%c_{n}  &  \simeq\frac{K}{n^{\gamma+\lambda}}%\frac{\left(  n!\right)
%^{\gamma+2\lambda}}{\left(  M_{\gamma+\lambda}\right)  ^{2n-1}}\simeq\frac
%{K}{n^{\gamma+\lambda}}\left(  \sqrt{2\pi n}\right)  ^{\gamma+2\lambda}%
%\exp\left(  \left(  \gamma+2\lambda\right)  n\log\left(  n\right)  -\left(
%\gamma+2\lambda\right)  n -\left(  2 %n-1\right)  \log\left(  M_{\gamma+\lambda
%}\right)  \right) \\
%\end{align*}}

Some careful computations combined with the use of Euler--Maclaurin formulas as well as Stirling's approximation for the factorial in \eqref{C1} then lead to desired asymptotics for the concentrations $c_n$ (cf.\eqref{C7}). We refer to the appendix \ref{appendix} for further details. 
\begin{comment}
\textcolor{red}{A: I will double check later how to summarize in a consistent way the following part ...... Do we want to put all the computations in the appendix?}
\end{comment}

%The flux of monomers from the region where the cluster sizes is of order one,
%to large sizes (specifically cluster sizes $n$ satisfying $n\gg\left(
%M_{\gamma+\lambda}\right)  ^{\frac{2}{\gamma+2\lambda}}$) is given by%
%\begin{equation}
%J\left(  M_{\gamma+\lambda}\right)  \simeq\left(  2\pi\right)  ^{??}%
%KM_{\gamma+\lambda}\exp\left(  -a\left(  M_{\gamma+\lambda}\right)  ^{\frac
%{2}{\gamma+2\lambda}}\right)  \label{MonFlux}%
%\end{equation}
%where%
%\[
%K=???\ \ ,\ \ a=\int_{0}^{\infty}\log\left(  1+\frac{1}{y^{\gamma+2\lambda}%
%}\right)  dy %\]

\bigskip

\bigskip 

\subsection{Solutions to \eqref{B6} for $M_{\gamma+\lambda} $ of order $1$}
\label{sec: inner behaviour when M constant} 
 In this section, we discuss formal asymptotics on the behaviour of the solutions to equation \eqref{B6}.
In particular, assuming that a solution to \eqref{B6} exists, we derive the following large-size behaviour for the solution $\overline{f}$ to equation \eqref{B6}
\begin{equation} \label{asympt of f when gamma+2lambda=1}
\overline{f}(x)\approx x^{\lambda-1} x^{- M_{\gamma+\lambda}^2 } \text{ as } x \rightarrow \infty
\end{equation} 
when $M_{\gamma +\lambda} \neq 1$ and 
\begin{equation} \label{asympt of f when gamma+2lambda=1, M=1}
\overline{f}(x) \approx \frac{1}{x ^{(\gamma + 3) /2 }  (\ln (x))^2 } \text{ as } x \rightarrow \infty
\end{equation} 
when $M_{\gamma+\lambda}=1$.

Notice that, when $M_{\gamma+\lambda} < 1$ the assumption that a solution $\overline f$ of \eqref{B6} exists together with \eqref{asympt of f when gamma+2lambda=1}, implies that
\[
\int_0^\infty x^{1 - \lambda} \overline{f}(x) dx = \infty. 
\]
This is a contradiction. Indeed a solution to equation \eqref{B6} should be such that $M_{1-\lambda} (\overline {f}) \leq 1/M_{\gamma+\lambda} < \infty $ as will be explained in Step 1. 
This is the reason why we expect solutions to equation \eqref{B6} to exist only when $M_{\gamma+\lambda} > 1$.

Before showing how we obtain the asymptotics \eqref{asympt of f when gamma+2lambda=1} and \eqref{asympt of f when gamma+2lambda=1, M=1} we stress that the behaviour for $\overline{f}$ in \eqref{asympt of f when gamma+2lambda=1, M=1} reminds the one derived in \cite{ferreira2019stationary} for $M_{\gamma+\lambda}=0$ and for $\gamma+2\lambda <1$, i.e., $\overline f \approx x^{- (\gamma +3) /2 } $ for $x $ large, but here we have logarithmic corrections.

The rest of the section is devoted to the derivation of \eqref{asympt of f when gamma+2lambda=1} under the assumption that $M_{\gamma+\lambda} \neq 1$. We do not show \eqref{asympt of f when gamma+2lambda=1, M=1} as it can be proven using very similar arguments. 
We present here preliminary results: 
a detailed and rigorous analysis of the existence of solutions for equation \eqref{B6} is challenging and goes beyond the aim of this paper.

We divide the derivation of \eqref{asympt of f when gamma+2lambda=1} in steps as follows. 

\bigskip

\textbf{Step 1: Reformulation of equation \eqref{B6}}  

\bigskip

As a first step to derive \eqref{asympt of f when gamma+2lambda=1}, it is convenient to rewrite equation \eqref{B6} in flux form as follows:  
\begin{equation} \label{flux formulation} 
J[\overline{f}](x)=\int_0^x z \eta(z) dz - M_{\gamma+\lambda} \int_0^x z^{1-\lambda} \overline{f}(z) dz. 
\end{equation}
Here $J[\overline{f}](x)$ is the flux of mass from the region of particles of size less than $x$ to the region of particles of sizes bigger than $x$ and is defined as in \eqref{eq:flux}, namely
\[
J[\overline{f}](x)=\int_0^x \int_{x-z}^\infty K(y,z) z \overline{f}(z) \overline{f}(y) dy dz.  
\]

Motivated by \eqref{flux formulation}, we define the function $g$ as $g(x):= x^{1-\lambda} \overline{f}(x) $. 
Then $g$ satisfies the following equation 
\begin{equation} \label{eq g} 
\tilde{J}[g](x)= \int_0^x z \eta(z) dz - M_{\gamma+\lambda} \int_0^x  g(z) dz
\end{equation}
where
\[
\tilde{J}[g](x):= \int_0^x \int_{x-z}^\infty W(y,z) g(z) g(y) dy dz=J[\overline{f}](x) \quad \text{ with } \quad W(y,z)=z^\lambda y^{\lambda-1}K(y,z). 
\]
We now study the behaviour of the solution $g$ to equation \eqref{eq g} as $x \rightarrow \infty$.

To this end we first show that 
\begin{align} \label{source and Mgamma+lambda}  
\int_0^\infty z \eta(z) dz = M_{\gamma+\lambda} \int_0^\infty  g(z) dz. 
\end{align} 
\bigskip

\textbf{Step 2: \eqref{source and Mgamma+lambda} holds}  

\bigskip

Since $g$ is a solution to equation \eqref{eq g}, then $\int_0^\infty g(x) dx \leq \frac{1}{M_{\gamma+\lambda}}. $
Assume, by contradiction, that $\int_0^\infty z \eta(z) dz \neq  M_{\gamma+\lambda} \int_0^\infty  g(z) dz$. 
This would imply that there exists a constant $c>0$ such that 
\[
\int_0^\infty z \eta(z) dz -  M_{\gamma+\lambda} \int_0^\infty  g(z) dz=c= \lim_{x \rightarrow \infty } \tilde{J}[g](x). 
\]
Since in \cite{ferreira2019stationary} it has been proven that the equation for the constant flux solution $J[f]=c$ does not have solutions when $\gamma+2\lambda \geq 1$, we deduce that $c=0.$ 
Therefore, \eqref{source and Mgamma+lambda} follows. 

\bigskip

\textbf{Step 3: equation for $\int_x^\infty g(y) dy$ } 

\bigskip

As a consequence of Step 2, considering large values of $x $ in equation \eqref{eq g}, we have that 
\begin{align}\label{simplified eq g}
\tilde{J}[g](x)=M_{\gamma+\lambda} \int_x^\infty g(z) dz.
\end{align} 
We introduce the notation $G(x):= \int_x^\infty g(y) dy $. Thanks to Step 2 we know that $G(x) \rightarrow 0$ as $x \rightarrow \infty$. 
Moreover \eqref{source and Mgamma+lambda} implies that 
\[
G(x) \int_0^x g(z) dz  =\frac{1}{M_{\gamma+\lambda} } G(x) - G(x)^2
\]
where we normalized the source $\eta$ so that $\int_0^\infty x \eta (x) dx =1 $. 

The above observation allows to rewrite equation \eqref{simplified eq g} as 
\begin{equation} \label{eq for approx G}
\int_0^x g(z) \int_x^\infty \left[ W(y,z) -1 \right]g(y) dy dz + \int_0^x g(z) \int_{x-z}^x W(y,z) g(y) dy dz =\left( M_{\gamma+\lambda}  - \frac{1}{M_{\gamma+\lambda} }\right) G(x) + G(x)^2. 
\end{equation} 
We recall that, since the source is zero in $(0,1]$, then $\overline{f}((0,1))=0$ and $g((0,1))=0$. 

\bigskip

\textbf{Step 4: the first term in left-hand side of \eqref{eq for approx G} tends to zero faster than $G(x) $ as $x \rightarrow \infty$}

\bigskip 

\begin{comment}
    One should not use $\delta(x)$ as the notation here: soon it will denote the Dirac delta.
\end{comment}

Let us consider a function $\varepsilon(x)$ such that $\varepsilon =\varepsilon(x) \rightarrow 0$ as $x \rightarrow \infty $ so slowly that $G(\varepsilon(x) x ) \rightarrow 0$ as $x \rightarrow \infty$. 
We can rewrite the first term on the left hand side of equation \eqref{eq for approx G} as follows 
\begin{align*} 
& \int_0^x g(z) \int_x^\infty \left[ W(y,z) -1 \right]g(y) dy dz = \int_1^{\varepsilon x} g(z)  \int_x^\infty \left[ W(y,z) -1 \right]g(y) dy dz \\
&+ \int_{\varepsilon x}^x g(z) \int_x^\infty \left[ W(y,z) -1 \right]g(y) dy dz . 
\end{align*} 
 Using the fact that on the domain of integration of the first integral on the right-hand side of the equality above we have that $z \ll y $,
we can use the approximation for $K$ given by \eqref{S1E8}, together with the fact that $\gamma+2\lambda=1$, to deduce that for large values of $x$
\[
\int_1^{\varepsilon x} g(z) \int_x^\infty \left[ W(y,z) -1 \right]g(y) dy dz  \leq  \varepsilon G(1) G(x)
\]
where $\varepsilon=\varepsilon (x)$ tends to zero as $x \rightarrow \infty$.

On the other side, using again the properties of the kernel $W$ we deduce that
\begin{align*} 
 \int_{\varepsilon x}^x g(z) \int_x^\infty \left[ W(y,z) -1 \right]g(y) dy dz  \leq G(\varepsilon x ) G(x). 
\end{align*} 
The desired conclusion then follows. 

\bigskip

\textbf{Step 5: approximation of the second term in left-hand side of \eqref{eq for approx G} as $x \rightarrow \infty$}

\bigskip

We rewrite the second term in the left-hand side of equation \eqref{eq for approx G} as follows 
\begin{align*}
& \iint_{S_x } W(y,z) g(y) g(z) dy dz = \iint_{ S_x \cap \{  y < \varepsilon z \} }  W(y,z) g(y) g(z) dy dz  + \iint_{ S_x \cap \{ \varepsilon z < y < \frac{z}{\varepsilon}  \} }    W(y,z) g(y) g(z)  dy dz \\
&+ \iint_{ S_x \cap \{y > \frac{z}{\varepsilon}  \} }    W(y,z) g(y) g(z)  dy dz 
\end{align*} 
where $S_x:=\{ (y,z) \in \mathbb R_* \times \mathbb R_* : 0<z<x, \ x-z < y <x\} $.

Now notice that
\begin{align*}
 \iint_{ S_x \cap \{ \varepsilon z < y < \frac{z}{\varepsilon}  \} }    W(y,z) g(y) g(z)  dy dz \leq (G(\varepsilon x ))^2. 
\end{align*} 
On the other side, using the asymptotics of the kernel $W$, we deduce that, as $x \rightarrow \infty$
\begin{align*}
& \iint_{ S_x \cap \{  y < \varepsilon z \} }  W(y,z) g(y) g(z) dy dz
+ \iint_{ S_x \cap \{y > \frac{z}{\varepsilon}  \} }    W(y,z) g(y) g(z)  dy dz \\
&\approx \int_1^{\varepsilon x } \int_{x-z}^x \left( 1 + \frac{y}{z } \right) g(y) g(z) dy dz  \approx x \int_1^{\varepsilon x } \frac{G(x-z) - G(x) }{z }  g(z) dz \approx - x G(1)  \frac{ d G(x) }{ d x }. 
\end{align*}

\bigskip

\textbf{Step 6: asymptotics of $f$}

\bigskip

Using the asymptotics computed in Steps 4 and 5 as well as the fact that $M_{\gamma+\lambda}>1$ and keeping only the leading terms in \eqref{eq for approx G}, we deduce 
\[
- x  G(1)\frac{d G(x) }{ dx } \approx \left( M_{\gamma+\lambda} - \frac{1}{M_{\gamma+\lambda} } \right) G(x) 
\]
as $x$ tends to infinity.
Therefore, for large sizes $x$ we expect that 
\[
G(x) \approx  x^{- (M_{\gamma+\lambda}^2 -1 ) },
\]
where we used \eqref{relation in and out moments} since $G(1)=M^{in}_{1-\lambda}$. 
From this we infer that, for large sizes, $g(x) \approx x^{- M_{\gamma+\lambda}^2 } $ and, therefore, \eqref{asympt of f when gamma+2lambda=1} follows.

\section{Anomalous self-similarity for $\gamma+2\lambda>1$}\label{sec:anomalous}
In this section we study the anomalous self-similar solutions to equation \eqref{eq:coag with growth} under the assumptions that $\gamma+2\lambda >1 $ and $\gamma\leq -1$.
Notice that in all the cases specified below we expect that these self-similar solutions describe the long-time behaviour of equation \eqref{S1E1}. The self-similar behaviour need to be matched with \eqref{MonFlux} in the region of small sizes. 

We divide this section according to different assumptions on the parameters $\gamma$ and $\lambda $ that lead to different scalings. 
 We start briefly recalling the dimensional analysis in Section \ref{sec:dimensional} and summarizing the self-similar behaviour derived in this section, in the regime $\gamma+2\lambda > 1 $, $\gamma\leq -1$.

\begin{itemize}
    \item If  $ \gamma <-1$ and $\gamma+\lambda>0$,
    the dimensional analysis in Section \ref{sec:dimensional} implies that
$M_{0}$ approaches to a constant as $t\rightarrow\infty$. This is due to the fact that  the rate of
destruction of clusters due to the coagulation process is slow and at the same time the influx of small particles is negligible, more precisely since $\left[  M_{\gamma
+\lambda}\right]  \approx t^{\gamma+\lambda}$ then $J_c\left(  M_{\gamma
+\lambda}\right)  $ is exponentially small. 
    The scaling is 
    \[ 
    L(t) = t .
    \] 
    The self-similar solution is of the form given by a Dirac $\delta$-function,
    $$f(t,x)=\frac{b}{t} \delta\left( \frac{x}{t} - \frac{1}{b}\right) . $$ Here, the constant $b$ depends on the initial data.
\item If $\gamma<-1$, %\gamma \leq -1$ 
    and $\gamma+\lambda \leq 0 $, the dimensional analysis made in Section \ref{sec:dimensional} suggests that $M_0$ tends to infinity as time tends to infinity. 
This is due to the influx of small particles in the system which, in this case, is stronger than the coagulation term. For these reasons, we expect the moment $M_{\gamma+\lambda}$ to increase at most logarithmically in time so that $J_c(M_{\gamma+\lambda})$ is not negligible.
     The scaling is 
     \[ 
     L(t) = \left(\frac{t}{M_{\gamma+\lambda}} \right)^{\frac{1}{1-\gamma-\lambda}}.
     \] 
    The corresponding self-similar solution will be 
    \begin{equation*}
       %\label{eq:selfsim_2} 
       f(t,x) = \frac{t}{\left(\frac{t}{M_{\gamma+\lambda}}\right)^{\frac{2}{1-(\gamma+\lambda)}}} \Phi_\infty \left(\frac{x}{\left( \frac{t}{M_{\gamma+\lambda}}\right)^{\frac{1}{1-(\gamma+\lambda)}}} \right)
    \end{equation*}
    where $\Phi_\infty $ is given by
    \[
\Phi_\infty(\xi)= \frac{C}{\xi^{\gamma+\lambda} } \exp\left(\int_0^\xi
\frac{1}{\left( 1-(\gamma+\lambda) - \eta^{1-(\gamma+\lambda)} \right) } \frac{d \eta }{\eta^{\gamma+\lambda}}\right).
\]
    \item If $\gamma=-1$ (hence $\gamma+\lambda>0$),
    the dimensional analysis made in Section \ref{sec:dimensional} suggests that $M_0$ tends to zero as time goes to infinity. This is due to the fact that the contribution to the evolution of the number of particles of the coagulation operator is larger than the influx of particles $[J_c(M_{\gamma+\lambda})]$. 
Indeed, in this case we have that $M_{\gamma+\lambda} \sim t^{\lambda -1} ( \log (t))^{1/2 (\lambda -2)} $, hence, $M_{\gamma+\lambda}$ increases like a power law. 
The contribution due to the injected small particles to the total number of particles, $J_c(M_{\gamma+\lambda}) $, is therefore negligible.
The characteristic length is 
\[L(t) = t (\ln(t))^{1/2}.
\] 
The corresponding self-similar solution is $$f(t,x) = \frac{1}{ a t \ln(t)} \delta\left( \frac{x}{t \ln(t)^{1/2}} - a\right).$$ Here the constant $a$ depends on the kernel, i.e., $a^2=K(1,1)$.
\end{itemize}

\subsection{Self-similarity when $\gamma+\lambda >0$, $\gamma <-1$} 
\label{sec:gamma+lambda>0}

In Section \ref{sec:dimensional} we showed  that when $\gamma+\lambda > 0$, $\gamma < -1$ and $\gamma+2\lambda > 1$, equation \eqref{S1E1} is compatible with a self-similar behaviour. 
In particular, we expect the solutions $f$ to equation \eqref{S1E1} to behave, as time goes to infinity and in the region of large sizes, as self-similar solutions to equation \eqref{eq:coag with growth} with the following scaling properties: 
 \begin{equation} \label{scaling properties gamma+lambda>0} 
[x]= \left[ t \right]\quad \text{ and } \quad [f] = \frac{1}{[t]}. 
\end{equation}
The aim of this Section is to study these self-similar solutions and check that they are consistent with the scenario derived in Section \ref{sec:dimensional}. 
\subsubsection*{Time dependent equation in self-similar variables}
Motivated by \eqref{scaling properties gamma+lambda>0}, we consider the change of variables 
\[
f(t,x)=\frac{1}{t} \Phi(\tau, y) \quad  \text{ with } \ \tau=\ln(t) \ \text{ and } \ y= \frac{x}{t}
\]
in equation \eqref{eq:coag with growth} and deduce that the function $\Phi$ should satisfy the following equation 
\begin{equation}\label{eq:Phi_eq}
\partial_\tau \Phi(\tau,y)  - \Phi(\tau, y) - y \partial_y \Phi (\tau, y) +\frac{ \partial_y (y^{\gamma+\lambda} \Phi(\tau,y) ) }{\int_{\mathbb R_*} x^{\gamma+\lambda } \Phi(\tau,x ) dx } =e^{(1+\gamma) \tau } \mathbb K [\Phi](\tau, y)
\end{equation}
We recall from the dimensional analysis of Section \ref{sec:dimensional} that the moment $M_0[\Phi](\tau)$  converges to a constant as time goes to infinity. 
 We will denote this constant by $M_{0,\infty} := \lim_{\tau\to \infty} M_0[\Phi](\tau)$.

Since we are assuming that $1+\gamma <0 $, the coagulation term in the above equation becomes negligible as $\tau$ tends to infinity. 
Hence, the above equation can be approximated by the following equation,  for large $\tau$, 
\begin{equation}\label{eq:approx PDE gamma+lambda>0}
\partial_\tau \Phi(\tau,y)  +\partial_y \left(  \left( \frac{ y^{\gamma+\lambda}  }{M_{\gamma+\lambda} } - y \right)  \Phi(\tau,y)\right) =0.
\end{equation}

We make the %self-similar 
ansatz that $\Phi(\tau,\cdot) \to \Phi_b$ as $\tau \to \infty$, where $\Phi_b$ is a solution to
\begin{equation}\label{eq:stat_approx_gamma+lambda>0}
\frac{d}{dy} \left(  \left( \frac{ y^{\gamma+\lambda}  }{M_{\gamma+\lambda} } - y \right)  \Phi(y)\right) =0
\end{equation}
 We will see in the rest of this subsection that this ansatz is self-consistent. 

\subsubsection*{The self-similar profile} 
As explained in more detail below, \eqref{eq:stat_approx_gamma+lambda>0} has a family of solutions 
 $\Phi_b$ given by 
\begin{equation}\label{eq:family_Dirac}
\Phi_b(y)= b \,  \delta\left(y - \frac{1}{b}\right), \quad b>0.
\end{equation}
To compute the value of $b$,
we impose that 
$\int_0^\infty \Phi_b(y)dy  =  M_{0,\infty}.$
 Hence,
 \begin{equation} \label{eq:b}
 b =  M_{0,\infty}.
 \end{equation}
We note that the value of $b$ depends on the initial value $\Phi(y,0)$ because $M_0$ is not constant under the evolution \eqref{eq:approx PDE gamma+lambda>0}.
Actually, $M_0$ is modified  significantly for times $\tau$ of order $1$ by the coagulation term on the right hand-side of \eqref{eq:Phi_eq}
 in a way which strongly  depends on the initial data.
 On the other hand, we will check next that $\Phi(y,\tau)$ can approach $\Phi_b(y)$ in \eqref{eq:family_Dirac}, with $b>0$ arbitrary. We stress that this is very different from the situation that takes place in the critical case $\gamma=-1, \gamma+2\lambda >1$ (see Section \ref{sec:gamma=-1}).
 For this critical case, a suitable function $\Phi (y,\tau)$ approaches to  $b  \delta(y-\frac{1}{b})$ as $\tau\to \infty$, but the value of $b$ is uniquely determined by the collision kernel $K$ independently of the initial data.

\subsubsection*{Convergence to the self-similar profile} 
We will now show that the steady states $\Phi_b$ defined in \eqref{eq:family_Dirac} are asymptotically stable as $\tau \to \infty$ by using the method of characteristics to solve \eqref{eq:approx PDE gamma+lambda>0}.

%\bigskip

%\subsection{Change of variables}

%\subsection{Solution of \eqref{eq:approx PDE gamma+lambda>0} via the equations for characteristics}

We start by applying a change of variables to obtain a simpler form of the model.
 We define the new  variables 
$$G(w,s) = y^{\gamma+\lambda} \Phi(y,\tau),\quad  k  w = \frac{y^{1-(\gamma+\lambda)}}{1-(\gamma+\lambda)},\quad s= (1-(\gamma+\lambda))\tau,$$
where the constant $k$ is defined by
$$\frac{
k^a}{(1-(\gamma+\lambda))^{\frac{1}{1-(\gamma+\lambda)}}} = 1, \quad \text{ with } \quad a := \frac{\gamma+\lambda}{1-(\gamma+\lambda)}.$$

Then, \eqref{eq:approx PDE gamma+lambda>0} becomes
\begin{equation}
\frac{\partial G}{\partial s} + \frac{\partial}{\partial w} \left( \left( \frac{1}{\int_0^\infty G(\eta,s)\eta^a d\eta }- w \right)G \right) = 0,\quad w>0,\ s >0
\end{equation}
with boundary conditions 
\begin{equation}
G(0,s) = 0,\ s >0,\quad G(w,0) = G_0(w),\ w>0
\end{equation}
where the initial value is given by $G_0(w) = y^{\gamma+\lambda}\Phi(y,0)$.

The equations for characteristics are given by 
\begin{eqnarray}\label{eq:cccc}
\frac{\partial w}{\partial s}(s, w_0) = - w(s,w_0) + h(s), \quad \frac{dG}{d s} (w(s,w_0),s)=0, \quad w(0,w_0) = w_0
\end{eqnarray}
where 
$$h(s) := \frac{1}{\int_0^\infty G(\eta,s)\eta^a d\eta}.$$

Notice that we can describe the evolution of $G$ using the solutions to the characteristics $w$. For $w_0>0$ and $s \geq 0$, $G$ is given by 
$$G(w(s,w_0),s) = \frac{G_0(w_0)}{\frac{\partial w}{\partial w_0}(s,w_0)}, \quad w(0,w_0) = w_0.$$
Notice also that \eqref{eq:cccc} implies  
$$\frac{\partial }{\partial s} \left( \frac{\partial w}{\partial w_0}\right) = - \frac{\partial w}{\partial w_0}, \quad \frac{\partial w}{\partial w_0}(0,w_0) = 1. $$
Then 
\begin{equation}\label{eq:partial_w}
 \frac{\partial w}{\partial w_0} (s, w_0) = e^{-s}.
 \end{equation}

\begin{figure}[h]%
\centering
\includegraphics[width=0.5\textwidth]{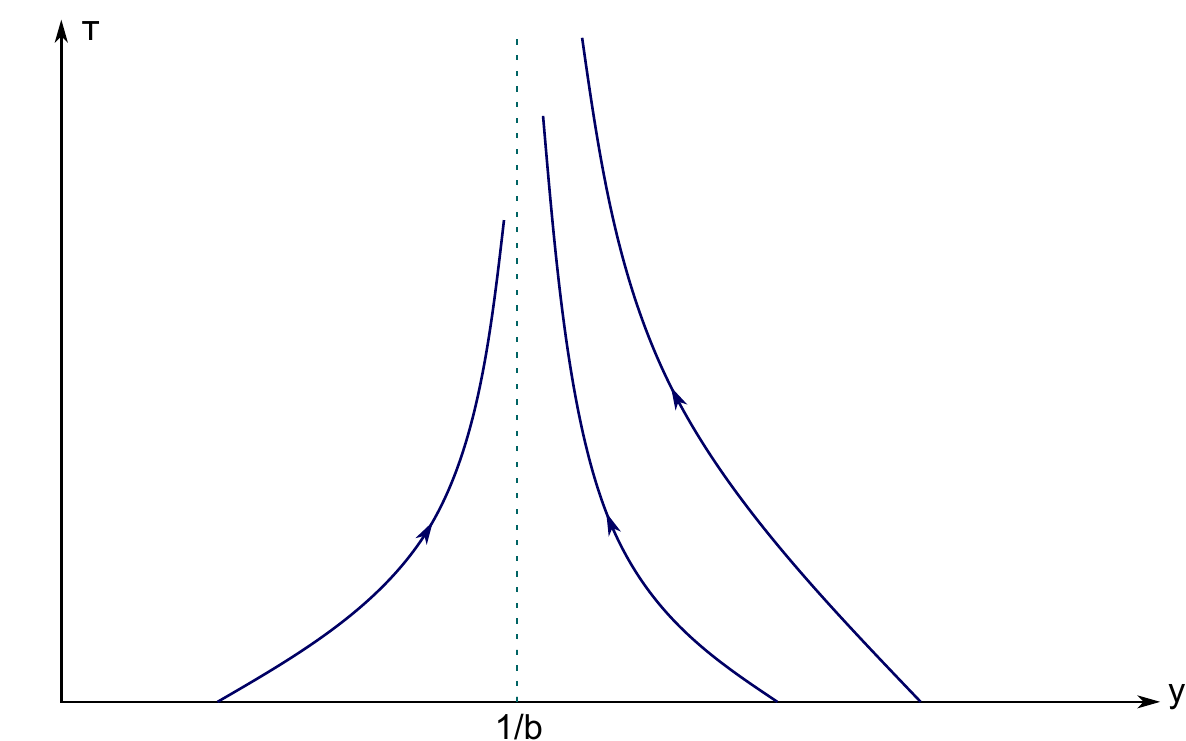}
\caption{Structure of the characteristic curves.}
\label{fig4}
\end{figure}

Suppose that most of the initial mass $G_0$ is  concentrated in a region $[\varepsilon_0, 1/\varepsilon_0]$, and that
$$\int_0^\infty G_0(w)(1+w^a)dw < \infty.$$
 Then, due to \eqref{eq:partial_w}, the length of the interval containing  most of the mass shrinks exponentially. More precisely, the length at time $s$ is $(\frac{1}{\varepsilon_0} - \varepsilon_0)e^{-s}$  
and one can  prove that $G(w,s)$ converges to a Dirac mass supported at some $\bar w$ as $s \to \infty$, i.e., 
$$G(w,s) \to  m \delta \left(w-\bar w\right)  \text{ as }  s \to \infty$$ 
for some $m>0$. % where $m = \int_0^\infty G_0(w)dw$. 
See Figure \ref{fig4} for a visual representation. 
We next compute the values of $m$ and $\bar w$ in terms of $b$ and the kernel parameters. Since $G$ approaches to a Dirac, then  $h$ converges to a constant. %$m^{-1}(\bar w)^{-a}$.
From the equation for characteristics \eqref{eq:cccc} it follows that $w(s,w_0)$ converges to $m^{-\frac{1}{1+a}}$ as $s\to \infty$, therefore $\bar w = m^{-\frac{1}{1+a}}$ ensures that $h(\infty)=\overline w $. 
Using the change of variable we have that $M_0[\Phi](\tau) = k M_0[G](s)$, therefore, from \eqref{eq:b}, it follows that 
$\lim_{s \to \infty} M_0[G](s) = \frac{b}{k},$
which determines $m$ as $m = \frac{b}{k}$. Then
$\bar w = \left(\frac{b}{k}\right)^{-\frac{1}{1+a}}$. Hence $ \Phi(\tau, y) \rightarrow \Phi_b (y)$ as $\tau \rightarrow \infty$.  

\subsection{Self-similarity when $\gamma+\lambda \leq0$, $\gamma <-1$ }%and $\gamma+2\lambda >1 $}
\label{sec:gaamma+lambda<0}
In Section \ref{sec:dimensional} we showed  that when $\gamma+\lambda \leq 0$, $\gamma < -1$ and $\gamma+2\lambda > 1$, equation \eqref{S1E1} is compatible with a self-similar behaviour with the following scaling properties: 
 \begin{equation} \label{scaling properties gamma+lambda<0} 
[x]= \left[ \frac{t}{M_{\gamma+\lambda}} \right]^{\frac{1}{1-(\gamma+\lambda) }}\quad \text{ and } \quad [f] = \frac{[t]}{\left[ \frac{t}{M_{\gamma+\lambda}} \right]^{\frac{2}{1-(\gamma+\lambda) }}}. 
\end{equation}
The aim of this Section is to study these self-similar solutions and check that they are consistent with the scenario derived in Section \ref{sec:dimensional}. 

\subsubsection*{Time dependent equation in self-similar variables} 
Since the natural scaling for the anomalous self-similar solutions are the ones in \eqref{scaling properties gamma+lambda<0}, we consider the following self-similar change of variables in equation \eqref{eq:coag with growth}
\begin{equation}\label{self-simi scaling gamma+lambda<0}
f(t, x)= \frac{t}{\left( \frac{t}{M_{\gamma+\lambda}} \right)^{\frac{2}{1-(\gamma+\lambda) }}} \Phi(\tau,\xi) \quad  \text{ with } \ \xi=\frac{x}{\left( \frac{t}{M_{\gamma+\lambda} }\right)^{\frac{1}{1-(\gamma+\lambda) }} } \ \text{ and } \ \tau=\ln(t).
\end{equation}
Substituting \eqref{self-simi scaling gamma+lambda<0} in equation \eqref{eq:coag with growth}, we deduce that the function $\Phi$ satisfies the following equation
\begin{align} \label{eq:time dep gamma+lambda<0}
& \partial_\tau \Phi(\tau, \xi) - \frac{1+(\gamma+\lambda)}{1-(\gamma+\lambda)} \Phi(\tau, \xi) - \frac{\xi \partial_\xi \Phi(\tau,\xi) }{1-(\gamma+\lambda)} +\frac{2}{1-(\gamma+\lambda)} \left(  \frac{  \partial_{\tau} M_{\gamma+\lambda} }{M_{\gamma+\lambda} }\right) \Phi(\tau,\xi) \\
& + \frac{1}{1-(\gamma+\lambda) } \left( \frac{\partial_\tau M_{\gamma+\lambda }}{M_{\gamma+\lambda}}\right) \xi \partial_\xi \Phi(\tau,\xi) + \partial_\xi \left( \xi^{\gamma+\lambda} \Phi(\tau,\xi )\right) = e^{- \frac{\gamma+2\lambda -1}{1-(\gamma+\lambda)} \tau} \left( M_{\gamma+\lambda} \right)^{\frac{1-\gamma}{1-(\gamma+\lambda)}} (\mathbb K \Phi) (\tau,\xi) \nonumber 
\end{align}
for $\tau>0$ and $\xi >0$. 

Notice that, since $\gamma+2\lambda > 1 $, the contribution to the coagulation operator term is negligible compared to the growth term. This is in agreement with the dimensional considerations performed in Section \ref{sec:dimensional} and with the fact that $M_0 \rightarrow \infty $ as time goes to infinity. As explained in Section \ref{sec:dimensional}, we expect $M_{\gamma+\lambda} $ to increase at most logarithmically, in particular as in \eqref{asymp of Mgamma+lambda}.
This implies that 
\begin{equation} \label{log decay of M}
\partial_\tau M_{\gamma+\lambda} = t \partial_t M_{\gamma+\lambda} \ll M_{\gamma+\lambda} \text{ as } \tau=\ln(t) \rightarrow \infty.
\end{equation} 
Moreover, since equation \eqref{eq:coag with growth} is compatible with the existence of self-similar solutions, we  make the %self-similar 
ansatz: we assume that $\Phi(\tau, \cdot) \rightarrow \Phi_\infty $ as $\tau \rightarrow \infty$. 
Using the fact that $\gamma+2\lambda > 1$, the self-similar ansatz and \eqref{log decay of M}, we deduce that the self-similar profile $\Phi_\infty$  is the solution to the following equation 
\begin{equation}\label{eq:self-sim profile gamma+lambda<0}
- \frac{ 1+\gamma+\lambda}{1-(\gamma+\lambda)} \Phi_\infty(\xi) - \frac{1}{1-(\gamma+\lambda)} \xi \partial_\xi \Phi_\infty(\xi) + \partial_\xi\left( \xi^{\gamma+\lambda} \Phi_\infty (\xi) \right) =0,\quad \text{ for } \xi \geq 0 . 
\end{equation}

 We now compute $\Phi_\infty$ and then we show that the self-similar ansatz 
 %assumption that the solutions $f$ to equation \eqref{S1E1} are such that
 %\[
%f(t,x) \simeq  \frac{t}{\left( \frac{t}{M_{\gamma+\lambda}} \right)^{\frac{2}{1-(\gamma+\lambda) }}} \Phi_\infty(\xi) \quad  \text{ with } \ \xi=\frac{x}{\left( \frac{t}{M_{\gamma+\lambda} }\right)^{\frac{1}{1-(\gamma+\lambda) }} } \ \text{ and } \ \tau=\ln(t), \quad \text{ as } \tau \rightarrow \infty \text{ and } \xi \rightarrow \infty,
%\]
is compatible with \eqref{log decay of M}.

\subsubsection*{The self-similar profile} \label{sec:ss profile gamma+lambda<0} 
The function $\Phi_\infty$ defined as 
\[
\Phi_\infty(\xi)= \frac{C}{\xi^{\gamma+\lambda} } \exp\left(\int_0^\xi
\frac{1}{\left( 1-(\gamma+\lambda) - \eta^{1-(\gamma+\lambda)} \right) } \frac{d \eta }{\eta^{\gamma+\lambda}}\right) 
\]
and such that $\Phi_\infty(\xi) =0$ for every $\xi > \xi^*$, with $\xi^*:= (1-(\gamma+\lambda))^{\frac{1}{1-(\gamma+\lambda)}}$, 
 satisfies \eqref{eq:self-sim profile gamma+lambda<0} in weak sense. Here $C$ is the normalization constant given by 
\[
\frac{1}{C}:=\int_0^{\xi^*} \frac{1}{\xi^{\gamma+\lambda} } \exp\left(\int_0^\xi
\frac{1}{\left( 1-(\gamma+\lambda) - \eta^{1-(\gamma+\lambda)} \right) } \frac{d \eta }{\eta^{\gamma+\lambda}}\right)  d\xi. 
\]
Therefore, $\Phi_\infty$ is the self-similar profile. 
 
%We rewrite equation \eqref{eq:self-sim profile gamma+lambda<0} as 
%\[
%\left( \xi^{\gamma+\lambda} - \frac{\xi}{1-(\gamma+\lambda) }\right) \partial_\xi \Phi_\infty ( \xi)=\left(  \frac{1+\gamma+\lambda }{1-(\gamma+\lambda)} - (\gamma+\lambda) \xi^{\gamma+\lambda -1} \right) \Phi_\infty ( \xi). 
%\]
%We remove the singularity at the origin with the following change of variables $ G(\xi) = \xi^{\gamma+\lambda} F_\infty (\xi)$. Indeed $G$ satisfies the following ODE
%\[
%\left(1- \frac{\xi^{1-(\gamma+\lambda)} }{1-(\gamma+\lambda)} \right) \partial_\xi G(\xi)  = \frac{1}{1-(\gamma+\lambda)}\frac{G(\xi)}{\xi^{\gamma+\lambda}}
%\]
%Let $\xi^*:=(1-(\gamma+\lambda))^{\frac{1}{1-(\gamma+\lambda)}}$. 
%When $\xi < \xi^* $, then $\xi^{\gamma+\lambda} - \frac{\xi}{1-(\gamma+\lambda) }>0$. 
%We deduce that
%\[
%G(\xi)= C  \exp\left(\int_0^\xi
%\frac{1}{\left( 1-(\gamma+\lambda) - \eta^{1-(\gamma+\lambda)} \right) } \frac{d \eta }{\eta^{\gamma+\lambda}}\right), \quad \text{ for }  0 < \xi < \xi^*
%\]
%for a positive constant $C$. 
%This implies that the function $\Phi_\infty$, defined as \[
%\Phi_\infty (\xi)=\frac{C}{\xi^{\gamma+\lambda} }  \exp\left(\int_0^\xi
%\frac{1}{\left( 1-(\gamma+\lambda) - \eta^{1-(\gamma+\lambda)} \right) } \frac{d \eta %}{\eta^{\gamma+\lambda}}\right), \quad \text{ for }  0 < \xi < \xi^*
%\] and such that $ \Phi_\infty(\xi)=0$ for every $\xi > \xi^*$, 
%where the constant $C$ is such that
%\[
%\int_{\mathbb R_*} \xi \Phi_\infty( \xi) d\xi= \int_0^{\xi^*} \xi \Phi_\infty( \xi) d\xi=1
%\] 
%satisfies equation \eqref{eq:self-sim profile gamma+lambda<0}

\subsubsection*{The properties of the self-similar profile are consistent with the ansatz \eqref{log decay of M}} \label{subsec:compatibility} 

Now we check that the properties of the self-similar profile $\Phi_\infty $ are consistent with \eqref{log decay of M}. 
To this end we use a matched asymptotics argument to combine the behaviour of $f$ in the inner region with the behaviour of $f$ in the outer region. 

It has been shown in Section \ref{sec:inner region} that, in the inner region, when $1 \ll n \ll L(t)=t^{\frac{1}{1-(\gamma+\lambda) }}$, the solution $f(t)$ to equation \eqref{eq:coag with growth} behaves as the function $c_n$ given by \eqref{C7}. Thanks to the Ansatz \eqref{log decay of M} and to the fact that $\gamma+2 \lambda > 1 $ we also have that $M_{\gamma+\lambda}^{\frac{2}{\gamma+2\lambda} } \ll t^{\frac{1}{1-(\gamma+\lambda) }} $ as $t \rightarrow \infty$. 

We can therefore consider the following regime:  $M_{\gamma+\lambda}^{\frac{2}{\gamma+2\lambda} } \ll n \ll t^{\frac{1}{1-(\gamma+\lambda) }} $ as $t \rightarrow \infty$, where we have that
\begin{equation}\label{c_n limit} 
c_n(t) \simeq \frac{\left[  \left(  2\pi\right)  ^{\frac{\gamma}{2}+\lambda
}K\right]  \left(  M_{\gamma+\lambda}\right)  ^{2}}{n^{\gamma+\lambda}%
}\exp\left(  -\left(  M_{\gamma+\lambda}\right)  ^{\frac{2}{\gamma+2\lambda}%
}A_{\gamma+2\lambda}\right)
\end{equation} 
as $t \rightarrow \infty$ and where 
\[
A_{\gamma+2\lambda} := \int_{0}^\infty\log\left(  1+\frac{1}{y^{\gamma+2\lambda}}\right)
dy. 
\]
On the other hand, when $n \approx t^{\frac{1}{1-(\gamma+\lambda) }} $ we expect the solution $f$ to equation \eqref{eq:coag with growth} to behave like  \eqref{self-simi scaling gamma+lambda<0}. Since we also have that $\Phi(\tau, \cdot) \rightarrow \Phi_\infty $ as $\tau$ tends to infinity, then 
\[
f(t,n)= t^{- \frac{1+\gamma+\lambda}{1-(\gamma+\lambda) }} M_{\gamma+\lambda}^{\frac{2}{1-(\gamma+\lambda)}} \Phi_\infty\left( \xi  \right) \quad \text{ with } \ \xi= n \left(\frac{M_{\gamma+\lambda}}{t}\right)^{\frac{1}{1-(\gamma+\lambda) }}. 
\]
Matching the inner region with the outer region and taking into account that $\Phi_\infty(\xi) \sim \xi^{-(\gamma+\lambda)}$ as $\xi \rightarrow 0^+$, we deduce that
\[
 \frac{\left[  \left(  2\pi\right)  ^{\frac{\gamma}{2}+\lambda
}K\right]  \left(  M_{\gamma+\lambda}\right)  ^{2}}{n^{\gamma+\lambda}%
}\exp\left(  -\left(  M_{\gamma+\lambda}\right)  ^{\frac{2}{\gamma+2\lambda}%
}A_{\gamma+2\lambda}\right)= t^{- \frac{1+\gamma+\lambda}{1-(\gamma+\lambda) }} M_{\gamma+\lambda}^{\frac{2}{1-(\gamma+\lambda)}}  n^{-(\gamma+\lambda)}\left(\frac{t}{M_{\gamma+\lambda}}\right)^{\frac{\gamma+\lambda}{1-(\gamma+\lambda) }}. 
\]
This implies that for a suitable constant $c>0$ we have that
\[
 c M_{\gamma+\lambda }^{- \frac{\gamma+\lambda}{1-(\gamma+\lambda) }} \exp\left( - M_{\gamma+\lambda}^{\frac{2}{\gamma+2\lambda}}A_{\gamma+2\lambda} \right) = t^{-\frac{1}{1-(\gamma+\lambda)}} 
\]
and hence, 
\[
M_{\gamma+\lambda}\simeq \left(\frac{1}{A_{\gamma+2\lambda}} \ln\left( c t^{\frac{1}{1-(\gamma+\lambda) }}\right) \right)^{\frac{\gamma+2\lambda}{2}} \text{ as } t \rightarrow \infty.
\]
This is in agreement with the Ansatz \eqref{log decay of M} as $t \rightarrow \infty$ and with \eqref{asymp of Mgamma+lambda}. 

\bigskip

\subsubsection*{Comparison with the physical literature} 
The result for $\gamma+2\lambda > 1$, $\gamma < -1$ and $\gamma+\lambda \leq 0$ obtained in this section is consistent with the results obtained in \cite{KMR98} and in \cite{KMR99}. 
In these references,   
the coagulation equation with a source of monomers with the kernel
\[
K_{ij} =i^{-\mu} + j^{-\mu}
\] 
has been considered.
This corresponds, in our notation, to $\gamma+\lambda=0$ and $\mu=\lambda=-\gamma> 1$. 
In these papers, it is also found that $c_n $ behaves as \eqref{c_n limit} with $M_0=M_{\gamma+\lambda}$. 
Moreover, they obtain the following growth rate for the $0$-th moment 
\[
M_{\gamma+\lambda}=M_0 \simeq \left\{ \frac{1}{A_{\gamma+2\lambda}} \ln \left[ t \ln\left( \frac{\ln t }{A_{\gamma+2\lambda}}\right)^{1-3\lambda/2} \right] \right\}^{\lambda/2}
\]
as $t \rightarrow \infty$. This is in agreement with \eqref{asymp of Mgamma+lambda}.

\subsection{Self-similarity when $\gamma=-1$ } \label{sec:gamma=-1}
In Section \ref{sec:dimensional} we showed that when $\gamma= -1 $ and $\gamma+2\lambda >1 $, equation \eqref{S1E1} is compatible with a self-similar behaviour. 
In particular we expect the following scaling properties:
\begin{equation} \label{scalings gamma=-1}
[x]= [t] [\ln(t)]^{\frac{1}{2}} \quad \text{ and } \quad [f]=\frac{1}{[t \ln(t)]}.
\end{equation} 
The aim of this Section is to study the self-similar solutions for equation \eqref{eq:coag with growth} corresponding to the above scalings.

\subsubsection*{Time dependent equation in self-similar variables} 
The scaling properties \eqref{scalings gamma=-1} suggest to make the  following change of variables in equation \eqref{eq:coag with growth}
\[
f(t,x) =  \frac{ 1}{t \ln t } \Phi(\tau, y ) \quad \text{ with } \quad \tau=\ln t \ \text{ and } \ y= \frac{x}{t \ln(t)^{1/2}}.
\]
By elementary computations, we deduce that $\Phi$ satisfies the following equation 
\begin{equation} \label{time dep Phi gamma=-1}
\partial_\tau \Phi(\tau,y) + \partial_y\left[\left( \frac{y^{\gamma+\lambda}}{M_{\gamma+\lambda} } - y \right)  \Phi(\tau,y) \right] -\frac{1}{2 y \tau } \partial_y (y^2 \Phi(\tau, y))= \frac{1}{\tau} \mathbb K (\Phi) (\tau, y). 
\end{equation} 
Notice that, due to the factor $1/\tau $ in front of it, we expect the contribution of the coagulation operator to the dynamics to be weaker than the one of the growth term. 
Differently from what happens when $\gamma<-1$, the coagulation term  %$\gamma+\lambda >0$ and $\gamma+\lambda \leq 0$, this term
is not exponentially decreasing in time. This explains in particular why we have a different scenario compared to the one that we have when $\gamma+\lambda >0$ and $\gamma<-1$, although in both  cases we have that the number of particles tends to a constant as time tends to infinity and that the $M_{\gamma+\lambda}$ grows in time like a power law. 

We now make the self-similar ansatz, hence we assume that $\Phi(\tau, \cdot) \rightarrow \Phi_\infty$ as $\tau $ tends to infinity. 
Using the fact that some of the terms in equation \eqref{time dep Phi gamma=-1} decay in time as $1/\tau$ we take formally the limit as $\tau $ tends to infinity in equation \eqref{time dep Phi gamma=-1} to deduce that 
$\Phi_\infty $ should satisfy the following equation 
\begin{equation}\label{ss profile gamma0-1} 
\partial_y\left[\left( \frac{y^{\gamma+\lambda}}{M_{\gamma+\lambda} } - y \right)  \Phi_\infty(y) \right]= 0. 
\end{equation}
In the next section we study the form of $\Phi_\infty$. 

\subsubsection*{The self-similar profile} 
In this section we want to study the solutions to equation \eqref{ss profile gamma0-1}. 
Since we also want $\int_{\mathbb R_*} x \Phi_\infty(x) dx =1 $ we deduce that 
\begin{equation} \label{self sim profile gamma=-1} 
\Phi_\infty (y)=\frac{1}{a} \delta(y-a)
\end{equation}
for some $a >0$. 
We stress that this case is similar to the one studied in Section \ref{sec:gamma+lambda>0} (cf.\ equation \eqref{eq:stat_approx_gamma+lambda>0}). 
In that case, we compute the value of $b$ in \eqref{eq:b} using the fact that $M_0$ converges to a constant in time, implying that $b=M_{0,\infty}$ (see \eqref{eq:b}) and it hence depends on the 
zeroth moment of the initial datum. 

\begin{figure}[h]%
\centering
\includegraphics[width=0.6\textwidth]{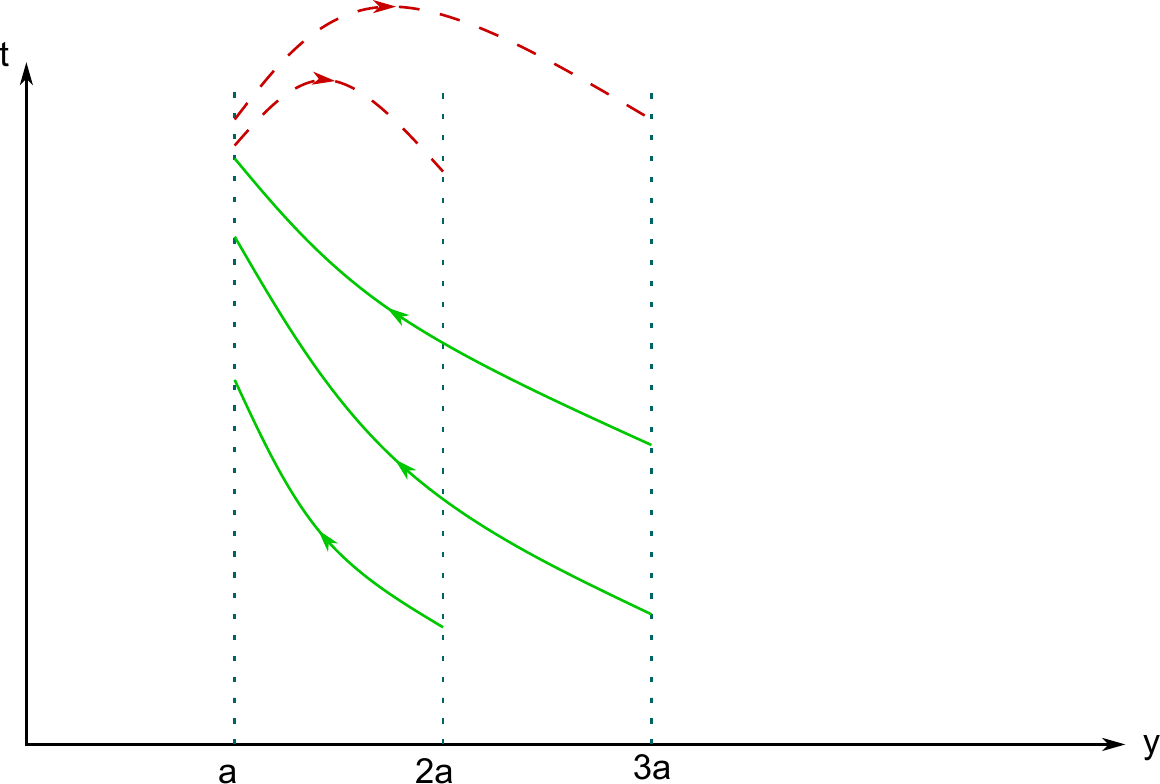}
\caption{In this figure, the solid line describes the characteristic curves transporting particles to size $y=a$. The dashed line, instead, represents the effect of the coagulation term.}
\label{fig5}
\end{figure}

The same type of argument does not work with the set of parameters considered in this section and we will have to adapt it to this case using the precise asymptotic behaviour of $M_0$. 
We therefore analyse in more detail the dynamics for large times.
A sketch of the interaction between the growth term and the coagulation term in equation \eqref{time dep Phi gamma=-1} is presented in Figure \ref{fig5}. As we will see later, even if the coagulation term tends to zero as time tends to infinity, it will determine the value of $a$ in \eqref{self sim profile gamma=-1}. 

To this end, using the fact that we expect $\Phi(\tau, \cdot) \rightarrow \frac{1}{a} \delta(\cdot-a)$ and using the homogeneity of the kernel $K$, we consider the following approximation for the coagulation operator 
\begin{align*}\label{approx for coagulation operator} 
& \mathbb K (\Phi)(\tau, y)  \simeq \frac{1}{2 a^2}\int_{0}%
^{y}K\left(  y-x,x\right)  \delta \left(y-x-a\right)  \delta\left( x- a\right)
dx- \frac{1}{a}\Phi \left(\tau,  y \right) \int_0^\infty K(x,y) \delta(x-a) dx 
\\
& \simeq  \frac{1}{2 a^3} K(1,1) \delta (y- 2 a) -\frac{1}{a^2} K(1,1) \Phi( \tau ,y).
\end{align*}

Using this approximation for the coagulation operator in equation \eqref{time dep Phi gamma=-1}, we deduce that
\begin{align*}
&\partial_\tau \Phi(\tau,y) + \partial_y\left[\left( \frac{y^{\gamma+\lambda}}{M_{\gamma+\lambda} } - y - \frac{1}{2 \tau } y \right)  \Phi(\tau,y) \right] = \frac{K(1,1)}{2 a^3 \tau } \delta(y- 2a) +  \frac{1}{2 \tau } \Phi(\tau, y) 
- \frac{1}{a^2\tau} K(1,1) \Phi(\tau, y)
\end{align*} 

Since the coagulation term in equation \eqref{time dep Phi gamma=-1} is of order $1/\tau$ and since the growth term, as time tends to infinity, is negative for sizes larger that $2a$, we can assume that $\Phi(\tau, \cdot)$ is equal to zero on the set $ (2a, \infty)$. 
As a consequence, we can reformulate the equation for $\Phi$ as 
\begin{equation}\label{approx eq gamma=-1} 
\partial_\tau \Phi(\tau,y) + \partial_y\left[\left( \frac{y^{\gamma+\lambda}}{M_{\gamma+\lambda} } - y - \frac{1}{2 \tau } y \right)  \Phi(\tau,y) \right] = \frac{1}{2 \tau } \Phi(\tau, y)- \frac{1}{a^2\tau} K(1,1) \Phi(\tau, y)\quad \text{ for } y < 2a 
\end{equation}
with the following boundary condition at $y=2a$ 
\begin{equation} \label{boundary condition at y=2a} 
- \left[ \frac{(2a)^{\gamma+\lambda}}{M_{\gamma+\lambda} } - 2a - \frac{a}{\tau}  \right] \Phi(\tau, 2a) = \frac{K(1,1) } {2a^3 \tau}, 
\end{equation}
which is obtained by integrating the above equation over $(2a,\infty)$.
 
Integrating equation \eqref{approx eq gamma=-1} from $0$ to $2a $, using the boundary condition \eqref{boundary condition at y=2a} and using the fact that $M_0$ approaches to a constant as time goes to infinity, we deduce that we should have 
\[
\left( \frac{K(1,1)}{a^2 \tau }- \frac{1}{2\tau }\right) \int_0^\infty \Phi(\tau,y) d y = \frac{K(1,1)}{2 a^3 \tau }.
\]
Using again the approximation $\Phi(\tau, \cdot) \simeq \frac{1}{a}\delta( \cdot-a) $, and hence the fact that $\int_{\mathbb R_*} \Phi(\tau, x) dx  \simeq \frac{1}{a} $, we deduce that
\begin{equation}\label{value a} 
a^2 =K(1,1).
\end{equation} 
Equation \eqref{approx eq gamma=-1} then reduces to 
\begin{equation}\label{approx eq gamma=-1 K=a2}  
\partial_\tau \Phi(\tau,y) +\left( \frac{y^{\gamma+\lambda}}{M_{\gamma+\lambda} } - y - \frac{y}{2 \tau }  \right)  \partial_y \Phi(\tau,y) + \left[  (\gamma+\lambda) \frac{y^{\gamma+\lambda-1}}{M_{\gamma+\lambda}} -1 \right] \Phi(\tau,y) = 0 \quad \text{ for } y <2a 
\end{equation}
with the following boundary condition at $y=2a$
\begin{equation} \label{boundary condition at y=2a K=a2} 
- \left[ \frac{(2a)^{\gamma+\lambda}}{M_{\gamma+\lambda} } - 2a - \frac{a}{\tau}  \right] \Phi(\tau, 2a) = \frac{1} {2a \tau}. 
\end{equation}
 
\subsubsection*{Convergence to the self-similar profile} 
We study equation \eqref{approx eq gamma=-1 K=a2} with the boundary condition \eqref{boundary condition at y=2a K=a2} and initial value $\Phi_0= \Phi(\tau_0, \cdot) $ at time $\tau = \tau_0 \ll 1$, supported in $y < 2a$. 
The families of characteristics for equation \eqref{approx eq gamma=-1 K=a2} are
\[
\frac{d Y}{ d \tau } = \frac{Y^{\gamma+\lambda} }{M_{\gamma+\lambda}} - Y - \frac{Y}{ 2 \tau}, \quad \text{ for } \tau \geq \tau_0 \ \text{ with }\  Y(\tau_0, y_0)=y_0 \in (0, 2a)
\]
and 
\[
\frac{d \overline Y}{ d \tau } = \frac{\overline Y^{\gamma+\lambda} }{M_{\gamma+\lambda}} - \overline Y - \frac{\overline Y}{ 2 \tau}, \quad \text{ for } \tau \geq \overline \tau \ \text{ with }\ \overline Y(\overline \tau, \overline \tau )=2a. 
\]
Here $\overline Y(\tau, \overline \tau )= Y(\tau- \overline \tau, 2a)$ where $\bar \tau>\tau_0 $. 
Hence, the solution $\Phi$ to equation \eqref{approx eq gamma=-1 K=a2} satisfies 
\begin{align*}
\Phi(\tau, Y(\tau, y_0))= \Phi_0(y_0) e^{\int_{\tau_0}^\tau \left( 1- (\gamma+\lambda) \frac{Y(s, y_0) ^{\gamma +\lambda -1}}{M_{\gamma+\lambda} }\right)  ds }
\end{align*} 
and using the boundary condition \eqref{boundary condition at y=2a K=a2} we obtain
\begin{align}\label{Phi boundary} 
\Phi(\tau, \overline Y(\tau, \overline \tau ))= \Phi(\overline \tau, 2a ) e^{\int_{\overline \tau}^\tau \left( 1- (\gamma+\lambda) \frac{ \overline Y(s, \overline \tau ) ^{\gamma +\lambda-1}}{M_{\gamma+\lambda} }\right)  ds }= \frac{1}{ 2 a^2 \overline \tau } \frac{1}{(2-2^{\gamma+\lambda} + \frac{1}{\overline \tau} )} e^{\int_{\overline \tau}^\tau \left( 1- (\gamma+\lambda) \frac{ \overline Y(s, \overline \tau ) ^{\gamma +\lambda-1 }}{M_{\gamma+\lambda} }\right)  ds }. 
\end{align} 

The contribution of the initial datum $\Phi_0$ as time tends to infinity is negligible.  Indeed, it is possible to show that 
\[
\int_{(0, 2a)} Y(\tau, y_0) \Phi(\tau, Y(\tau, y_0)) d y_0 \rightarrow 0 \text{ as }  \tau \rightarrow \infty. 
\]
Therefore, the asymptotic behaviour of $\Phi$ is given by the long-time behaviour of \eqref{Phi boundary}. 
To study the long-time behaviour of $\overline Y$, we approximate the equation for the characteristics as 
\[
\frac{d \overline Y} {d \tau } = \frac{\overline Y^{\gamma+\lambda} }{a^{\gamma+\lambda-1}} - \overline Y , \quad \text{ for } \tau \geq \overline \tau \ \text{ with }\ \overline Y(\overline \tau, \overline \tau )=2a
\]
where we are using the fact that as $\tau - \overline \tau \rightarrow \infty $ we have $M_{\gamma+\lambda} \simeq a^{\gamma+\lambda-1}$ because we expect $\Phi(\tau, \cdot) $ to converge to $\frac{1}{a} \delta_a$. 
Notice that as $\tau- \overline \tau  \rightarrow \infty$ we have that 
$ \overline Y(\tau, \overline \tau ) \rightarrow a$. 

Linearizing the ODE  for $\overline Y$ around the equilibrium $a$, we deduce that as $\tau-\overline \tau \rightarrow \infty $ we have that 
\[
\overline Y(\tau, \overline \tau) = Y(\tau- \overline \tau , 2a) \simeq a+ K e^{-[1-(\gamma+\lambda) ](\tau-\overline \tau)}, 
\] 
where the constant $K \simeq a$ depends on $\gamma+\lambda $ and on $a$. 
This implies that if $\xi=\overline Y(\tau, \overline \tau)$, then, since $a<\xi<2a$ and  $0< \frac{\xi-a}{K} < 1$, we have that
\[
\overline \tau= \tau+ \frac{1}{1-(\gamma+\lambda) } \ln \left( \frac{\xi-a}{K} \right) \leq \tau. 
\]

As a consequence, equality \eqref{Phi boundary} can be rewritten as 
\begin{align} \label{approx Phi gamma=-1}
\Phi(\tau, \xi) \simeq \frac{1}{ 2 a^2 \overline \tau } \frac{1}{(2-2^{\gamma+\lambda} + \frac{1}{\overline \tau} )} e^{( 1- \gamma-\lambda) (\tau-\overline \tau)  } \simeq 
\frac{1}{ 2 a^2 \overline \tau } \frac{1}{(2-2^{\gamma+\lambda}  )} \frac{K}{\xi-a}
\end{align} 
for large values of $\tau$ and $\overline \tau$ and for $\xi >a$. 
This is compatible with $\Phi(\tau, \cdot) \rightarrow \Phi_\infty$ as $ \tau \rightarrow \infty$ only when $\overline \tau \approx \tau$. 
Instead when $\overline \tau \ll \tau $, in particular, if 
\[
\tau \approx \frac{1}{1-(\gamma+\lambda)} \left| \log \left( \frac{\xi-a}{K} \right) \right| \,,
\]
the approximation is not valid. Indeed, according to \eqref{approx Phi gamma=-1}, we would have that $M_0 $ tends to infinity as $\tau $ tends to infinity. 

This argument partially supports the convergence towards the self-similar profile \eqref{self sim profile gamma=-1} where $a^2=K(1,1)$. However, a complete result would require more detailed analysis.

%\newpage

\section{Conjectures on the self-similar behaviour when $\gamma+2\lambda=1$} \label{sec:gamma=-1 and gamma+2lambda=1}
\begin{figure}[h!]%
\centering
\includegraphics[width=1.1\textwidth]{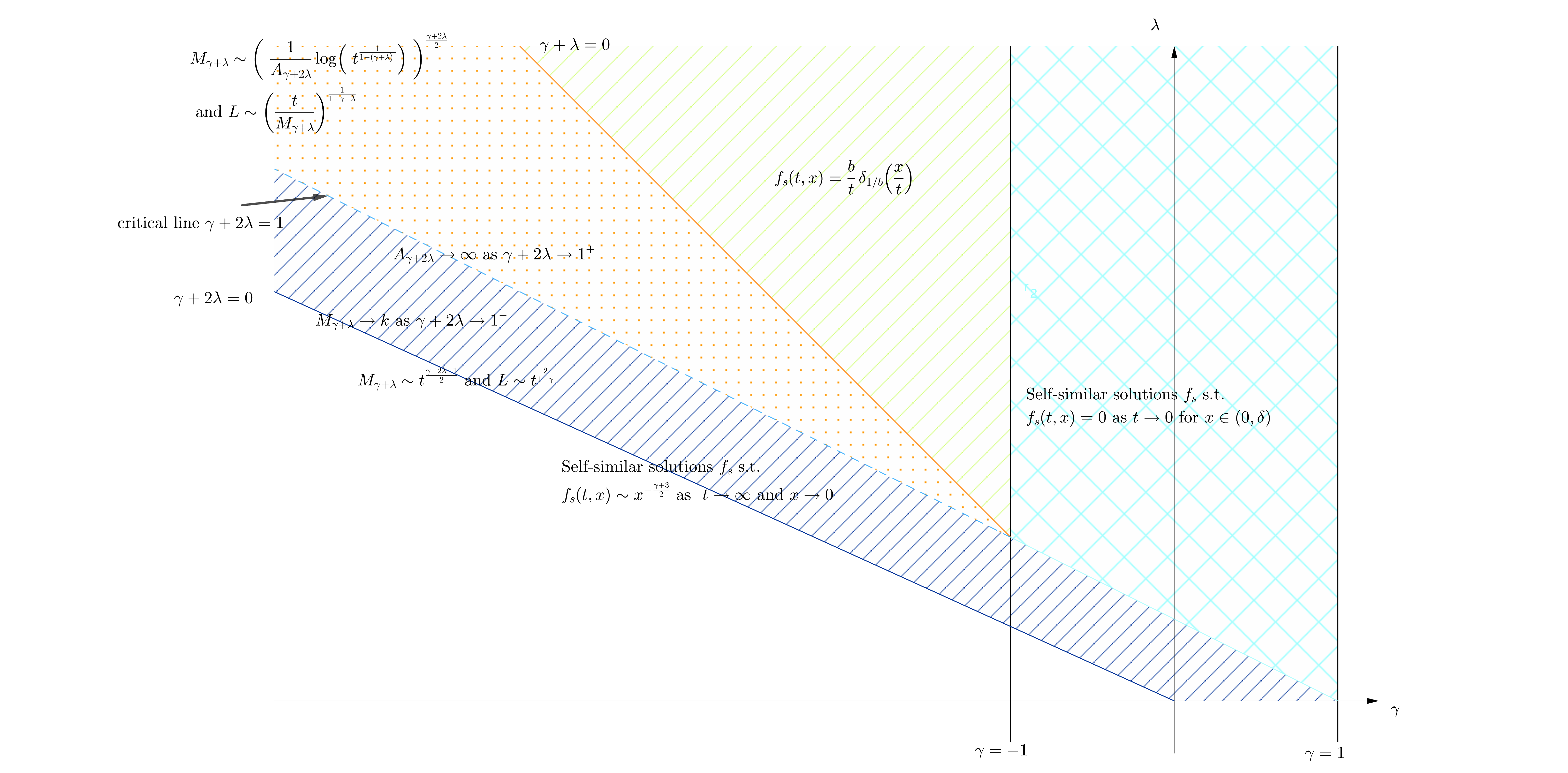}
\caption{In this picture we show the different behaviours near the critical line $\gamma+2\lambda=1$ in the different regimes $\gamma<-1$, $\gamma=-1$ and $-1 < \gamma < 1$.}
\label{fig6}
\end{figure}

As explained in the previous  Sections, when $\gamma+2\lambda >1 $, we obtained consistent asymptotic behaviour both for $\gamma>-1$, $\gamma=-1$ and $\gamma <-1$ with $M_{\gamma+\lambda } \rightarrow \infty $ as time tends to infinity. 
Instead, as explained in \cite{cristian2022long}, in the case $\gamma >-1$, $\gamma+2\lambda =1 $ there exist self-similar solutions for which the moment $M_{\gamma+\lambda}$ is constant. 
In the rest of this section, we first recall in Section  \ref{sec:comparison}  the results obtained in \cite{cristian2022long} for the case $\gamma>-1$, $\gamma+2\lambda =1$ and  describe then in  Section \ref{sec:matching} a possible scenario for the behaviour of the solutions when $\gamma \leq -1 $ and $\gamma+2\lambda=1 $. 
We stress that this case is more involved than the other as the matching between the inner and outer regions is non-trivial, see Section \ref{sec:matching} for more details. 
Figure \ref{fig6} also shows that the critical line $\gamma+2\lambda=1$ is at the intersection of a number of different behaviours.  There is no obvious choice between them for a dominant behaviour at the critical line which makes this case particularly challenging. 

\subsection{Comparison with previous results in the case $\gamma>-1$}\label{sec:comparison}

\subsubsection*{Self-similarity when $\gamma+2\lambda =1 $ and $\gamma >-1$}
 The long-time behaviour of the solutions $f $ to equation \eqref{eq:coag with growth} when $\gamma+2\lambda=1 $ and $\gamma> -1 $ is studied in \cite{cristian2022long}. 
In particular, it is explained there, via formal arguments, that the solutions $f$ approach, as time tends to infinity, the solution $\overline f $ to equation \eqref{B6} in the inner region where $x \approx 1 $. 
On the other hand, the dimensional analysis in Section \ref{sec:dimensional} shows that we can expect the long-time behaviour of the solutions to equation \eqref{eq:coag with growth} to be self similar and with the scalings given by \eqref{standard self similar solution}. The self-similar profiles $\Phi$, then satisfy the following equation
\begin{align} \label{eq standard self simi variables}
    -\frac{3+\gamma}{1-\gamma}\Phi\left(  \xi\right)  -\frac{2}{1-\gamma}\xi \partial_\xi
\Phi\left(  \xi\right)  +\frac{1}{M_{\gamma+\lambda}} \frac{\partial}{\partial\xi
}\left(  \xi^{\gamma+\lambda}\Phi\left(  \xi\right)  \right)  =\mathbb{K}%
\left[  \Phi\right]  \left(  \xi\right)  \ \ ,\ \ \xi>0.
\end{align}

In \cite{cristian2022long} the existence of these self-similar solutions is proven and their properties are analysed. 
In particular, it is proven in there that the self-similar profiles are zero in an interval near the origin and that they decay exponentially for large sizes. 
While the exponential decay for large sizes is typically seen in the theory of self-similar solutions for coagulation equations, the fact that the self-similar profile is zero near zero is very peculiar to the coagulation equation with source and to the assumption $\gamma+2\lambda \geq 1 $. 

In the rest of the section, we analyse the possibilities of having self-similar solutions of the form \eqref{standard self similar solution} with $\Phi$ satisfying \eqref{eq standard self simi variables} when $\gamma+2\lambda =1 $, $\gamma \leq -1$.

\subsubsection*{The argument applied for $\gamma+2\lambda >1 $ and $\gamma+\lambda \leq 0$ breaks down as $\gamma+2\lambda \rightarrow 1^+$} 
 The analysis done in Section \ref{sec:gaamma+lambda<0} for $\gamma+2\lambda > 1 $ and $\gamma+\lambda \leq 0$ shows that in that case the self-similar profile $\Phi_\infty $ is such that $\Phi_\infty(\xi) \sim \xi^{- (\gamma+\lambda)} $ as $\xi \rightarrow 0.$
Since this depends only on the sign of $\gamma+\lambda $, taking the limit as $\gamma+2\lambda \rightarrow 1^+$, we would expect the same behaviour for the self-similar profile corresponding to the case $\gamma+2\lambda=1$ and $\gamma \leq -1 $ (assuming that this self-similar profile exists).

We recall that the dimensional analysis made in Section \ref{sec:dimensional} for $\gamma+2\lambda >1$ and $\gamma+\lambda \leq 0$, gives the characteristic length \eqref{L char} and implies that $M_{\gamma+\lambda}$ is logarithmically increasing in time.
The dimensional analysis is valid also when $\gamma +2\lambda =1 $ and $\gamma \leq -1$. 
Therefore, if a self-similar solution exists it should have the following scalings
\[
x = \xi \left(\frac{t}{ M_{\gamma+\lambda}}\right)^{\frac{1}{1-(\gamma+\lambda)}} \quad f(t,x) = \frac{t}{\left( \frac{t}{M_{\gamma+\lambda}} \right)^{\frac{2}{1-\gamma-\lambda}}} \Phi(\tau, \xi) . 
\] 
The function $\Phi$, then, should satisfy the following equation 
\begin{align} \label{eq:time dep gamma+2lambda=1 wrong}
& \partial_\tau \Phi(\tau, \xi) - \frac{1+(\gamma+\lambda)}{1-(\gamma+\lambda)} \Phi(\tau, \xi) - \frac{\xi \partial_\xi \Phi(\tau,\xi) }{1-(\gamma+\lambda)} +\frac{2}{1-(\gamma+\lambda)} \left(  \frac{  \partial_{\tau} M_{\gamma+\lambda} }{M_{\gamma+\lambda} }\right) \Phi(\tau,\xi) \\
& + \frac{1}{1-(\gamma+\lambda) } \left( \frac{\partial_\tau M_{\gamma+\lambda }}{M_{\gamma+\lambda}}\right) \xi \partial_\xi \Phi(\tau,\xi) + \partial_\xi \left( \xi^{\gamma+\lambda} \Phi(\tau,\xi )\right) =  \left( M_{\gamma+\lambda} \right)^{\frac{1-\gamma}{1-(\gamma+\lambda)}} (\mathbb K \Phi) (\tau,\xi) \nonumber 
\end{align}
for $\tau>0$ and $\xi >0$. 
The dominant term in equation \eqref{eq:time dep gamma+2lambda=1 wrong} would be the coagulation term, in contrast with the scenarios described in Section \ref{sec:anomalous}  for  $\gamma+2\lambda>1$.

On the other hand, the asymptotics derived in Section \ref{sec:inner region gamma+lambda to infinity} for the case $\gamma+2\lambda >1 $ 
%and $\gamma >-1 $
is based on the fact that,  when $\gamma+2\lambda \geq 1$, the flux of particles $J_c(M_{\gamma+\lambda} )$ is given by \eqref{C7}. 
When $\gamma+2\lambda =1 $, this reduces to 
\[
J_c(M_{\gamma+\lambda} )(x)= \frac{\left[  \left(  2\pi\right)  ^{\frac{1}{2}
}K\right]  \left(  M_{\gamma+\lambda}\right)  ^{2}}{x^{\gamma+\lambda} \sqrt{1+\frac{\left(  M_{\gamma+\lambda}\right)  ^{2}}{x}}
}\exp\left(  -\left(  M_{\gamma+\lambda}\right)  ^{2} %
 \int_0^{\frac{x}{(M_{\gamma+\lambda})^{2}}} \log\left( 1+\frac{1}{y}\right) d y \right). 
\]

In order to match the inner behaviour with the outer self-similar behaviour, we consider in  Section  \ref{sec:gaamma+lambda<0} that  $ M_{\gamma+\lambda } \ll x \ll t^{\frac{1}{1-(\gamma+\lambda) }}$, as $t\to \infty$ in the formula for the flux of particles. 
When $\gamma+2\lambda >1 $, we have that 
\[
 \int_0^{\frac{x}{(M_{\gamma+\lambda})^{\frac{2}{\gamma+2\lambda }}}} \log\left( 1+\frac{1}{y^{\gamma+2\lambda}}\right) d y \rightarrow A_{\gamma+2\lambda }
\]
as $\frac{x}{(M_{\gamma+\lambda})^{\frac{2}{\gamma+2\lambda }}} \rightarrow \infty$, 
where $A_{\gamma+2\lambda}:=\int_0^\infty  \log\left( 1+\frac{1}{y^{\gamma+2\lambda}}\right) d y  < \infty $. 
Then, for large $x$, the flux of particles behaves like \[
J_c(M_{\gamma+\lambda}) \approx  \left(  2\pi\right)  ^{\frac{\gamma+2\lambda}{2}}
K  \left(  M_{\gamma+\lambda}\right)  ^{2} \exp\left(  -\left(  M_{\gamma+\lambda}\right)  ^{2} %
A_{\gamma+2\lambda} \right) .
 \] 
 However, the constant $A_{\gamma+2\lambda} $ is equal to $\infty $  when $\gamma+2\lambda =1 $. 
As a consequence the matching argument explained in Section \ref{sec:gaamma+lambda<0} breaks down in that case. The case $\gamma+2\lambda =1 $ and $\gamma=-1 $ has already been considered in \cite{KMR98} and \cite{KMR99}, where a logarithmic correction for $M_{\gamma+\lambda} $ is suggested.

\subsection{Conjecture on the existence of self-similar solutions 
%to equation \eqref{S1E1}
when $M_{\gamma+\lambda} < \infty$}
%{The assumption $M_{\gamma+\lambda} < \infty$ is compatible with the existence of self-similar solutions of equation \eqref{S1E1}} 
\label{sec:matching}
In this section, we analyse the possibility of having self-similar solutions when $M_{\gamma+\lambda} $ is constant. 
When $\gamma+2\lambda=1 $ and $\gamma \leq -1 $, we could expect, as explained in Section \ref{sec:regions},  the solutions to equation \eqref{S1E1} to behave in the inner region ($x \approx 1$) as the solution $\overline f$ to \eqref{B6}, and, in the outer region ($x \approx t^{\frac{2}{1-\gamma} }$), as self-similar solutions to equation \eqref{eq:coag with growth} of the form \eqref{standard self similar solution}. We check here that this scenario does not lead to any contradictions. However, showing that the long-time behaviour to equation \eqref{S1E1} is indeed self-similar also when $\gamma+2\lambda =1$ and $\gamma \leq -1 $ would need a careful analytic study of the existence of self-similar solutions which we are not going to do here.

\subsubsection*{Matching between the inner and outer region}  
Since we are assuming that $M_{\gamma+\lambda } < \infty$, we deduce that the behaviour in the inner region is described by the solution $\overline f $ to equation \eqref{B6}. 
The asymptotics as $x $ tends to infinity of $\overline f $
suggest that if $M_{\gamma+\lambda}^{in} \geq  1$ then
\begin{equation} \label{eq limit gamma+2lambda<1}
\overline f (x) = x^{- \frac{\gamma+3}{2}} \varepsilon (x) 
\end{equation}
where $\varepsilon (x) \rightarrow 0$ as $x \rightarrow \infty$. 

In the outer region, when $x \approx t^{\frac{2}{1-\gamma}} $, we expect the solution $f$ to equation \eqref{S1E1} to behave, for large times, as the self-similar solutions of the form \eqref{standard self similar solution} where the self-similar profile $\Phi$ satisfies  equation \eqref{eq standard self simi variables}.

We now present asymptotic analysis suggesting that the self-similar profile may behave in two different ways, 
when $x \ll t^{\frac{2}{1-\gamma}}$, 
depending on whether $M_{\gamma+\lambda}=1$ or $M_{\gamma+\lambda} \neq 1$. Precisely, we expect that if $M_{\gamma+\lambda} =1$, then
\begin{equation} \label{condition at the origin}
\Phi(\xi) \approx \frac{1}{\xi^{\frac{\gamma+3}{2}} \left(\log (\frac{1}{\xi}) \right)^2}, \quad \xi \rightarrow 0
\end{equation}
 while if $M_{\gamma+\lambda}\neq 1 $, then
\begin{equation} \label{condition at the origin M 
 not 1}
\Phi(\xi) \approx \frac{1}{\xi^{\frac{\gamma+3}{2}}} \frac{1}{\xi^{(M_{\gamma+\lambda}^2 -1 )} },\quad \xi \rightarrow 0.
\end{equation}
 
 Although, if $\Phi $ behaves like \eqref{condition at the origin M 
 not 1} and $M_{\gamma+\lambda}>1$, 
 then we would have that the $\gamma +\lambda$ moment of $\Phi$ would be equal to infinity, leading to a contradiction. Therefore, we conjecture $M_{\gamma+\lambda} \leq 1 $.
 Even if we also know that $M_{\gamma+\lambda}=\frac{1}{M_{\gamma+\lambda}^{in}}$ and that $M_{\gamma+\lambda}^{in}\geq 1$, we do not know how to determine the value of $M_{\gamma+\lambda} \leq 1 $. 
 This could be studied as an eigenvalue problem for equation \eqref{eq standard self simi variables} that we do not plan to solve here.

We now explain how we derive \eqref{condition at the origin} and \eqref{condition at the origin M 
 not 1}. 
Since we are assuming that 
$M_{\gamma+\lambda } < \infty $, then we must have $\Phi(\xi) < \xi^{- \frac{3+\gamma}{2}}$ for small $\xi$. 
Therefore, considering only the leading terms as $\xi \rightarrow 0$ in equation \eqref{eq standard self simi variables} we deduce the following approximate equation
\[
\frac{1}{M_{\gamma+\lambda}}
\frac{\partial}{\partial\xi
}\left(  \xi^{\gamma+\lambda}\Phi\left(  \xi\right)  \right)  =\mathbb{K}%
\left[  \Phi\right]  \left(  \xi\right) \text{ as } \xi \rightarrow 0. 
\]
We now multiply by $\xi$ and integrate from $0$ to $z$ all the terms in the above equation. 
Using also the fact that $\lim_{ z \rightarrow 0} J[\Phi] (z)=0$, we rewrite the approximate equation in the following mass flux form
\begin{equation} \label{eq: asympt near 0 ss}
J[\Phi] (z)= - \frac{1}{M_{\gamma+\lambda}}z^{\gamma+\lambda +1 } \Phi(z) + \frac{1}{M_{\gamma+\lambda}}\int_0^z x^{\gamma+\lambda} \Phi(x) dx.
\end{equation}
Let us define the function $H(z):=\int_0^z x^{1-\lambda } \Phi(x) dx $. 
We now approximate $J$ with similar arguments to the one presented in Section \ref{sec: inner behaviour when M constant}. Indeed, assuming that $z \ll 1 $, using the asymptotics \eqref{S1E8} for the kernel and using the fact that $M_{\gamma+\lambda} < \infty $, we deduce that
\begin{equation} \label{approx flux at zero}
J[\Phi] (z) \approx  \int_0^z \int_z^\infty x K(x,y) \Phi(y) \Phi(x) dy dx \text{ as } z \rightarrow 0. 
\end{equation}
Then, assuming that $M_{\gamma+\lambda} =1 $, equation \eqref{eq: asympt near 0 ss} can be rewritten as 
\[
 (H(z))^2  \approx z \frac{d H(z)}{dz}. 
\]
This implies that $H(z) \approx \frac{1}{\log \left(1/z \right) } $ as $z $ tends to zero and, hence by differentiation, we deduce \eqref{condition at the origin}. 

When, instead, $M_{\gamma+\lambda} <1 $, we can use \eqref{approx flux at zero} to rewrite equation \eqref{eq: asympt near 0 ss} in the following approximated form:
\[
(1-M_{\gamma+\lambda}^2) H(z) \approx z \frac{d H(z)}{dz}. 
\]
This implies that $H(z) \approx z^{1-M_{\gamma+\lambda}^2} $ and hence \eqref{condition at the origin M 
 not 1} follows.

Notice that \eqref{condition at the origin} and \eqref{condition at the origin M 
 not 1} with $M_{\gamma+\lambda} \leq 1$ are compatible with \eqref{eq limit gamma+2lambda<1} with $M_{\gamma+\lambda}^{in} \geq 1$. Therefore, this scenario suggests that, in the case of the critical exponents $\gamma+2\lambda =1 $ and $\gamma \leq-1$, the long term behaviour of the solution $f$ to equation \eqref{S1E1} could be given by the self-similar solutions of the form \eqref{standard self similar solution} for equation \eqref{eq:coag with growth}, assuming that these solutions exist.

\subsubsection*{Relation with the non-existence result in \cite{cristian2022long}}
We recall that in \cite{cristian2022long} it has been proven that, when $\gamma+2\lambda >1 $ and $\gamma >-1 $, self-similar solutions for equation \eqref{eq:coag with growth} exist and that the self-similar profile $\Phi$ is zero near the origin. On the other side it has been proven there that self-similar solutions of the form \eqref{standard self similar solution} to equation \eqref{eq:coag with growth}  with $\int_0^1 x^{-\lambda } \Phi(x) dx < \infty $ do not exist.
The conjecture we make here is not in contradiction with the non-existence result stated in \cite{cristian2022long}. 
Indeed, the condition near the origin, \eqref{condition at the origin} as well as \eqref{condition at the origin M 
 not 1}, guarantees that 
\begin{equation*} 
\int_0^1 \xi^{-\lambda } \Phi(\xi) d \xi = \infty
\end{equation*}
for any choice of $M_{\gamma+\lambda} \geq 1$. 

It is interesting to compare the behaviour described here for $\gamma+2\lambda =1 $ and $\gamma \leq -1 $ to the one $\gamma+ 2 \lambda < 1$, see \cite{ferreira2021self}. In that case, the existence of self-similar solutions is proven for the coagulation equation with a constant influx coming from the origin.
This condition is reflected in the behaviour of the self-similar profile near the origin, i.e., $ \Phi (\xi) \approx \xi^{- \frac{(\gamma+3)}{2}}$ as $\xi $ tends to zero. 
In the case considered here, with $\gamma+2\lambda =1 $ and $\gamma \leq -1 $, we obtain in \eqref{condition at the origin} and in \eqref{condition at the origin M 
 not 1}  the same power law behaviour but with corrections. These corrections are due  to the fact that there are no fluxes from the origin.

\section{Conclusions}

In this work, relying also on \cite{cristian2022long}, \cite{ferreira2021self}, we have described several examples of long-time asymptotics for the solutions to the Smoluchowski coagulation equation with source, i.e., to \eqref{S1E1}. This shows that the equation exhibits a very rich class of possible self-similar behaviours, depending only on the range of parameters $\gamma$ and $\lambda$ which characterize the coagulation kernel.   
We summarize here how the self-similar behaviour of the solutions  $f$ to equation \eqref{S1E1} changes when varying $\gamma$ and $\lambda$.  

\begin{itemize} 

\item When $\gamma+2\lambda <1 $ we expect the following self-similar behaviour of $f$ for large $x$ and large $t$ 
\[ 
 f(t,x) \sim t^{- \frac{3+\gamma}{1-\gamma}} \Phi \left( \frac{x}{t^{\frac{2}{1-\gamma}}} \right) \text{ and } \Phi(\xi) \sim \xi^{- \frac{\gamma+3}{2}} \text{ as } \xi \rightarrow 0. 
\]
Rigorous results on the existence of self-similar solutions as above can be found in \cite{ferreira2021self}. 
While numerical simulations, showing convergence to self-similarity, supported also by matching asymptotics arguments can be found in \cite{davies1999smoluchowski}. 
\item When $\gamma+2\lambda >1 $, we expect the following self-similar behaviour of $f$ for large $x$ and large $t$  
\begin{center}
\renewcommand\arraystretch{2.5}
\begin{tabular}{ |l|l|}
\hline 
parameters & self-similar solution \\ \hline 
 $\pmb{\gamma >-1}$  &
       $ 
        f(t,x) \sim t^{- \frac{3+\gamma}{1-\gamma}} \Phi \left( \frac{x}{t^{\frac{2}{1-\gamma}}} \right) \ \ \text{ and} \  \ \Phi(\xi)=0 \ \text{ for small } \ \xi 
        $ \\ \hline 
 $\pmb{\gamma =-1}$  
 &
      $f(t,x) \sim \frac{1}{a t \ln t } \delta \left( \frac{x}{ t (\ln(t))^{1/2}  } -a  \right) $ \\
    \hline 
    $\pmb{\gamma <-1, \,  \gamma+\lambda >0}$
&
 $f(t,x) \sim  \frac{t}{\left(\frac{t}{M_{\gamma+\lambda}}\right)^{\frac{2}{1-(\gamma+\lambda)}}} \Phi\left(x{\left( \frac{M_{\gamma+\lambda}}{t}\right)^{\frac{1}{1-(\gamma+\lambda)}}} \right) $,  $\Phi(\xi) \sim \xi^{-(\gamma+\lambda)}$ as $\xi \rightarrow 0$
     \\ 
    \hline 
   $\pmb{\gamma <-1,  \, \gamma+\lambda  \leq 0}$ &
    $f(t,x) \sim \frac{1}{b t } \delta \left( \frac{x}{ t } -b  \right) $ \\\hline
\end{tabular}
\end{center}
Rigorous results on the existence of self-similar solutions when $\gamma > -1$ can be found in \cite{cristian2022long}. 
The analysis of the complementary cases
%, marked in blue, 
is the main content of this paper.

\item When $\gamma+2\lambda =1 $ we expect the following self-similar behaviour for large values of $x$ and $t$ 

\begin{comment}
\begin{center}
\renewcommand\arraystretch{2.5}
\begin{tabular}{ |c |}
\hline 
  $\pmb{\gamma >-1} $   \\ [0.5ex] 
\hline
        $f(t,x) \sim t^{- \frac{3+\gamma}{1-\gamma}} \Phi \left( \frac{x}{t^{\frac{2}{1-\gamma}}} \right) $ \ \ and \ \  $\Phi(\xi)=0$ for small $\xi$\\
        \hline \hline 
  $\pmb{\gamma \leq -1} $  \\ [0.5ex] 
\hline
       $f(t,x) \sim t^{- \frac{3+\gamma}{1-\gamma}} \Phi \left( \frac{x}{t^{\frac{2}{1-\gamma}}} \right) $ with 
       $\Phi(\xi) \sim \frac{\xi^{- \frac{\gamma+3}{2}}}{\log\left( 1/\xi \right) }$
 or  $\Phi(\xi) \sim \xi^{- \frac{\gamma+3}{2}+1-M_{\gamma+\lambda}} $ as $\xi \rightarrow 0$ \\\hline
\end{tabular}
\end{center}
\end{comment}

\begin{center}
\renewcommand\arraystretch{2.5}
\begin{tabular}{|l|l|}
\hline 
parameters & self-similar solution \\ \hline 
  $\pmb{\gamma >-1} $    &
        $f(t,x) \sim t^{- \frac{3+\gamma}{1-\gamma}} \Phi \left( \frac{x}{t^{\frac{2}{1-\gamma}}} \right) $ \ \ and \ \  $\Phi(\xi)=0$ for small $\xi$\\
        \hline 
  $\pmb{\gamma \leq -1} $ &
       $f(t,x) \sim t^{- \frac{3+\gamma}{1-\gamma}} \Phi \left( \frac{x}{t^{\frac{2}{1-\gamma}}} \right) $ with 
       $\Phi(\xi) \sim \frac{\xi^{- \frac{\gamma+3}{2}}}{\log\left( 1/\xi \right) }$
 or  $\Phi(\xi) \sim \xi^{- \frac{\gamma+3}{2}+1-M_{\gamma+\lambda}} $ as $\xi \rightarrow 0$ \\ \hline
\end{tabular}
\end{center}

\end{itemize}

A rigorous analysis  for $\gamma+2\lambda =1$ of the existence/non-existence of standard self-similar solutions which are zero near zero
%and $\gamma > -1 $
can be found in \cite{cristian2022long}. 
Here, we have presented conjectures in the case $\gamma+2\lambda =1 $ and $\gamma \leq -1 $ and the solutions then have different behaviours near zero. 
%These are marked in blue in the above table. 
We conjecture that, under these assumptions on the parameters $\gamma$ and $\lambda$, the characteristic length for the self-similar solutions is $L(t)=t^{\frac{2}{1-\gamma}}$. 
Moreover, using asymptotics analysis, we prove that the behaviour of the self-similar profile $\Phi$ near the origin depends on the value of $M_{\gamma+\lambda}$. We claim that if $M_{\gamma+\lambda} = 1 $ then  $\Phi(\xi) \sim \frac{1}{\log\left( 1/\xi \right) }\xi^{- \frac{\gamma+3}{2}}$ as $\xi \rightarrow 0$, while if $M_{\gamma+\lambda} >1 $ then $\Phi(\xi) \sim \xi^{- \frac{\gamma+3}{2}} \xi^{1-M_{\gamma+\lambda}} $ as $\xi \rightarrow 0$. 
A rigorous analysis would be needed to prove this conjecture since our computations are not enough to determine the value of $M_{\gamma+\lambda}$. 

\bigskip

\appendix
\section{ Some details on the justification of \eqref{C7}
%, solution to \eqref{B6} in the inner region
(cf.\ Section \ref{sec:inner region gamma+lambda to infinity}). %to  \eqref{A8da}-\eqref{A8d}.
}\label{appendix}

We now clarify the asymptotics (\ref{C1}). We first recall that the sequence $\left\{  X_{n}\right\}  _{n\in
\mathbb{N}}$ satisfies (\ref{B62}). It is convenient to  perform the change of variables %(motivated by (\ref{C1}))%
$X_{n}=\left(  n!\right)  ^{\gamma+2\lambda}Y_{n}$ in  (\ref{B62}). Noticing that $X_{1}=Y_{1}=1$ we obtain
\begin{align*}
%\left(  n!\right)  ^{\gamma+2\lambda}Y_{n}  &  =\frac{\left(  n\right)
%^{\lambda}}{2}\sum_{j=1}^{n-1}K\left(  n-j,j\right)  \left(  \left(
%n-j\right)  !\right)  ^{\gamma+2\lambda}\left(  \left(  j\right)  !\right)
%^{\gamma+2\lambda}Y_{n-j}Y_{j}\\
Y_{n}  &  =\frac{\left(  n\right)  ^{\lambda}}{2}\sum_{j=1}^{n-1}K\left(
n-j,j\right)  \frac{\left(  \left(  n-j\right)  !\right)  ^{\gamma+2\lambda
}\left(  \left(  j\right)  !\right)  ^{\gamma+2\lambda}}{\left(  n!\right)
^{\gamma+2\lambda}}Y_{n-j}Y_{j}\\
&  =\left(  n\right)  ^{\gamma+2\lambda}\frac{K\left(  n-1,1\right)  \left(
\left(  n-1\right)  !\right)  ^{\gamma+2\lambda}}{n^{\gamma+\lambda}\left(
n!\right)  ^{\gamma+2\lambda}}Y_{n-1}Y_{1}+\sum_{j=2}^{n-2}K\left(
n-j,j\right)  \frac{\left(  \left(  n-j\right)  !\right)  ^{\gamma+2\lambda
}\left(  \left(  j\right)  !\right)  ^{\gamma+2\lambda}}{\left(  n!\right)
^{\gamma+2\lambda}}Y_{n-j}Y_{j}\\
&  =\frac{K\left(  n-1,1\right)  }{n^{\gamma+\lambda}}Y_{n-1}+\sum_{j=2}%
^{n-2}K\left(  n-j,j\right)  \frac{\left(  \left(  n-j\right)  !\right)
^{\gamma+2\lambda}\left(  \left(  j\right)  !\right)  ^{\gamma+2\lambda}%
}{\left(  n!\right)  ^{\gamma+2\lambda}}Y_{n-j}Y_{j} \ .
\end{align*}

We need to examine the asymptotic behaviour of $Y_{n}$ as $n\rightarrow \infty.$ We remark that we can have exponential, ``subfactorial", corrections in
$Y_{n}.$ Actually, the detailed asymptotics of $Y_{n}$ might depend on the
detailed asymptotics of the function $F \left(  s\right)$ in \eqref{S1E7} as $s\rightarrow0$
(or $s\rightarrow1$). We remark also that  at the leading order we have
$\frac{K\left(  n-1,1\right)  }{n^{\gamma+\lambda}}\simeq\left(  \frac{n-1}%
{n}\right)  ^{\gamma+\lambda}\simeq1-\frac{\left(  \gamma+\lambda\right)  }%
{n}$ as $n\rightarrow\infty.$ %(ok since \gamma+\lambda<1) 
We then obtain the approximation 
\begin{equation}\label{Y_n_app}
Y_{n}=\left(  1-\frac{\left(  \gamma+\lambda\right)  }{n}\right)  Y_{n-1}%
+\sum_{j=2}^{n-2}K\left(  n-j,j\right)  \frac{\left(  \left(  n-j\right)
!\right)  ^{\gamma+2\lambda}\left(  \left(  j\right)  !\right)  ^{\gamma
+2\lambda}}{\left(  n!\right)  ^{\gamma+2\lambda}}Y_{n-j}Y_{j}. 
\end{equation}
The first two terms in \eqref{Y_n_app} yield an asymptotic behaviour with the form 
\begin{equation}
Y_{n}\simeq\frac{K}{n^{\gamma+\lambda}}\text{ as }n\rightarrow\infty.\label{C3}%
\end{equation}
It remains to check if the terms in the sum $\sum_{j=2}^{n-2}\left[  \cdot
\cdot\cdot\right]$ in \eqref{Y_n_app} provide a negligible contribution.  To do this, we assume
that $Y_{n-j}$ and$\ Y_{j}$ can be approximated by power laws as
indicated above. We then examine the contribution of the factorial terms. %Using the symmetry of the sum we can restrict to the case $j<\frac{n}{2}$. 
We
expect to have as the largest term the one with $j=2$. This term can be bounded as
\begin{align*}
  K\left(  n-2,2\right)  \frac{\left(  \left(  n-2\right)  !\right)
^{\gamma+2\lambda}\left(  \left(  2\right)  !\right)  ^{\gamma+2\lambda}%
}{\left(  n!\right)  ^{\gamma+2\lambda}}Y_{n-2}Y_{2}   \leq C\left(  n\right)  ^{\gamma+\lambda}\left(  \frac{1}{n^{2}}\right)
^{\gamma+2\lambda}\frac{1}{n^{\gamma+\lambda}}\leq\frac{C}{n^{2\left(
\gamma+2\lambda\right)  }} . %
\end{align*}
The effect of this term can be estimated using that $Y_{n}%
-Y_{n-1}=-\frac{\left(  \gamma+\lambda\right)  }{n}Y_{n-1}+O\left(  \frac
{1}{n^{2\left(  \gamma+2\lambda\right)  }}\right)  .$   This shows that the term is of
the order $\frac{C}{n^{2\left(  \gamma+2\lambda\right)  -1}}$ that is much smaller
than the term $\frac{K}{n^{\gamma+\lambda}}$ as $n\rightarrow\infty.$ 
%It remains to show that 
It is possible to show that  the remaining terms in the sum yield a
contribution smaller than $\frac{C}{n^{2\left(  \gamma+2\lambda\right)  }}.$ 
It then follows that the asymptotics (\ref{C3}) (and hence \eqref{C1}) holds in the limit $n\rightarrow\infty$.

\smallskip

Arguing similarly, we now compute the asymptotics \eqref{C7}  of the solutions for large values of $n$. 
%, assuming a similar approximation, but assuming that we keep the term $K\left(1,n\right)  \left(  n\right)  ^{\lambda}c_{1}$ in the denominator above. 
The
approximation of \eqref{A8d} becomes
\[
c_{n}\simeq\frac{\left(  n\right)  ^{\lambda}}{\left[  K\left(  1,n\right)
\left(  n\right)  ^{\lambda}c_{1}+M_{\gamma+\lambda}\right]  }K\left(
n-1,1\right)  c_{n-1}c_{1}\ \ \text{for }n\text{ sufficiently large.}%
\]
Using the approximation of $c_{1}$ in 
(\ref{B5}) we obtain 
\begin{equation}
c_{n}\simeq\frac{\left(  n\right)  ^{\lambda}}{\left[  K\left(  1,n\right)
\left(  n\right)  ^{\lambda}c_{1}+M_{\gamma+\lambda}\right]  }\frac{K\left(
n-1,1\right)  c_{n-1}}{M_{\gamma+\lambda}}\ \ \text{for }n\text{ sufficiently
large} \label{B7}%
\end{equation}
Assuming that this approximation is valid for $n\geq L$ with $L$ sufficiently
large, we obtain the following approximation for $c_{n}$ 
\[
c_{n}\simeq c_{L}\prod_{j=L+1}^{n}\left(  \frac{\left(  j\right)  ^{\lambda
}K\left(  j-1,1\right)  }{\left[  K\left(  1,j\right)  \left(  j\right)
^{\lambda}+\left(  M_{\gamma+\lambda}\right)  ^{2}\right]  }\right)
\]
and using that $K\left(  j-1,1\right)  \simeq K\left(  j-1,1\right)
\simeq\left(  j-1\right)  ^{\gamma+\lambda}$ for $j$ large, we get
\begin{equation}
c_{n}\simeq c_{L}\prod_{j=L+1}^{n}\left(  \frac{\left(  j-1\right)
^{\gamma+\lambda}}{\left[  \left(  j\right)  ^{\gamma+\lambda}+\frac{\left(
M_{\gamma+\lambda}\right)  ^{2}}{\left(  j\right)  ^{\lambda}}\right]
}\right)  \label{B8}%
\end{equation}
Thus, we obtain that the approximation of the cluster concentrations $c_{n}$ given
by (\ref{B8}) is valid for the range of values of $n$ for which we can 
use (\ref{B7}) as an approximation of (\ref{A8d}). 

 We can use (\ref{B8}) to
compute the flux of clusters towards large values of $n.$ To this end, we compute the asymptotic behaviour of the right hand side of (\ref{B8}) for
large values of $M_{\gamma+\lambda}$ and $n$ that can be rewritten as
\begin{align*}
 \prod_{j=L+1}^{n}\left(  \frac{\left(  j-1\right)  ^{\gamma+\lambda}%
}{\left[  \left(  j\right)  ^{\gamma+\lambda}+\frac{\left(  M_{\gamma+\lambda
}\right)  ^{2}}{\left(  j\right)  ^{\lambda}}\right]  }\right)  %\\
%&  =\exp\left(  \sum_{j=L+1}^{n}\log\left(  \frac{\left(  j-1\right)
%^{\gamma+\lambda}}{\left[  \left(  j\right)  ^{\gamma+\lambda}+\frac{\left(
%M_{\gamma+\lambda}\right)  ^{2}}{\left(  j\right)  ^{\lambda}}\right]
%}\right)  \right) \\
%& 
=\exp\left(  \sum_{j=L+1}^{n}\log\left(  \frac{\left(  j-1\right)
^{\gamma+\lambda}}{\left(  j\right)  ^{\gamma+\lambda}}\right)  -\sum
_{j=L+1}^{n}\log\left(  1+\frac{\left(  M_{\gamma+\lambda}\right)  ^{2}%
}{\left(  j\right)  ^{\gamma+2\lambda}}\right)  \right).
\end{align*}
As
$M_{\gamma+\lambda}\rightarrow\infty$, we can approximate the second sum inside the exponential as %an integral 
\begin{align*}
%&
\sum_{j=L+1}^{n}\log\left(    1+\frac{\left(  M_{\gamma+\lambda
}\right)  ^{2}}{\left(  j\right)  ^{\gamma+2\lambda}} \right)
%\\& 
=\left(  M_{\gamma+\lambda}\right)  ^{\frac{2}{\gamma+2\lambda}}%
\sum_{j=L+1}^{n}\log\left(    1+\frac{1}{\left(  \frac{j}{\left(
M_{\gamma+\lambda}\right)  ^{\frac{2}{\gamma+2\lambda}}}\right)
^{\gamma+2\lambda}} \right)  \frac{1}{\left(  M_{\gamma+\lambda
}\right)  ^{\frac{2}{\gamma+2\lambda}}} . %\\
%&  \simeq\left(  M_{\gamma+\lambda}\right)  ^{\frac{2}{\gamma+2\lambda}}%
%\int_{\frac{L}{\left(  M_{\gamma+\lambda}\right)  ^{\frac{2}{\gamma+2\lambda}%
%}}}^{\frac{n}{\left(  M_{\gamma+\lambda}\right)  ^{\frac{2}{\gamma+2\lambda}}%
%}}\log\left(  1+\frac{1}{y^{\gamma+2\lambda}}\right)  dy
\end{align*}
Using the Euler-McLaurin formula 
$ \sum_{j=m+1}^{n}f\left(  j\right)  =\int_{m}^{n}f\left(  x\right)  dx+\frac
{1}{2}\left(  f\left(  n\right)  -f\left(  m\right)  \right)  +... $ 
 we get 
\begin{align}\label{McLaurin}
&  \sum_{j=L+1}^{n}\log\left(  1+\frac{1}{\left(  \frac{j}{\left(
M_{\gamma+\lambda}\right)  ^{\frac{2}{\gamma+2\lambda}}}\right)
^{\gamma+2\lambda}}\right) 
%\\& 
=\int_{L}^{n}\log\left(  1+\frac{1}{\left(  \frac{x}{\left(  M_{\gamma
+\lambda}\right)  ^{\frac{2}{\gamma+2\lambda}}}\right)  ^{\gamma+2\lambda}%
}\right)  dx+\nonumber \\
&  +\frac{1}{2}\left[  \log\left(  1+\frac{1}{\left(  \frac{n}{\left(
M_{\gamma+\lambda}\right)  ^{\frac{2}{\gamma+2\lambda}}}\right)
^{\gamma+2\lambda}}\right)  -\log\left(  1+\frac{1}{\left(  \frac{L}{\left(
M_{\gamma+\lambda}\right)  ^{\frac{2}{\gamma+2\lambda}}}\right)
^{\gamma+2\lambda}}\right)  \right]
\end{align}
The next order term in the Euler-Maclaurin formula is bounded by the derivatives of the logarithmic function which give terms of order $\frac
{1}{L},\ \frac{1}{n}$ or smaller terms. Therefore, these contributions tend to 
zero since $L\gg1.$  
However, we must keep the terms in \eqref{McLaurin} because they give
relevant contributions. We will then approximate them as follows 
\begin{align*}
&\sum_{j=L+1}^{n}\log\left(  1+\frac{\left(  M_{\gamma+\lambda}\right)  ^{2}%
}{\left(  j\right)  ^{\gamma+2\lambda}}\right)  \\
&=\int_{L}^{n}\log\left(
1+\frac{\left(  M_{\gamma+\lambda}\right)  ^{2}}{x^{\gamma+2\lambda}}\right)
dx+\frac{1}{2}\left(  \log\left(  1+\frac{\left(  M_{\gamma+\lambda}\right)
^{2}}{n^{\gamma+2\lambda}}\right)  -\log\left(  1+\frac{\left(  M_{\gamma
+\lambda}\right)  ^{2}}{L^{\gamma+2\lambda}}\right)  \right). 
\end{align*}
\begin{comment}
\bigskip\ \texttt{... It is convenient to estimate the remainder in this
formula in order to check that it really converges to zero. ...}
\end{comment}
Using this new formula in (\ref{B8}) we would have 
\begin{align*}
c_{n}  &  \simeq 
%c_{L}\left(  \frac{L}{n}\right)  ^{\gamma+\lambda}\exp\left(
%-\left(  M_{\gamma+\lambda}\right)  ^{\frac{2}{\gamma+2%\lambda}}\int_{\frac
%{L}{\left(  M_{\gamma+\lambda}\right)  ^{\frac{2}{\gamma+2%\lambda}}}}%
%^{\frac{n}{\left(  M_{\gamma+\lambda}\right)  ^{\frac{2}%{\gamma+2\lambda}}}%
%}\log\left(  1+\frac{1}{y^{\gamma+2\lambda}}\right)  %dy\right)  \cdot\\
%&  \cdot\exp\left(  -\frac{1}{2}\left(  \log\left(  %1+\frac{\left(
%M_{\gamma+\lambda}\right)  ^{2}}{n^{\gamma+2\lambda}}\right) % -\log\left(
%1+\frac{\left(  M_{\gamma+\lambda}\right)  ^{2}}{L^{\gamma+2%\lambda}}\right)
%\right)  \right) \\
%&  =
c_{L}\left(  \frac{L}{n}\right)  ^{\gamma+\lambda}\exp\left(  -\left(
M_{\gamma+\lambda}\right)  ^{\frac{2}{\gamma+2\lambda}}\int_{\frac{L}{\left(
M_{\gamma+\lambda}\right)  ^{\frac{2}{\gamma+2\lambda}}}}^{\frac{n}{\left(
M_{\gamma+\lambda}\right)  ^{\frac{2}{\gamma+2\lambda}}}}\log\left(
1+\frac{1}{y^{\gamma+2\lambda}}\right)  dy\right) \times\\
&  \quad\times\left(  1+\frac{\left(  M_{\gamma+\lambda}\right)  ^{2}}%
{L^{\gamma+2\lambda}}\right)  ^{\frac{1}{2}}\left(  1+\frac{\left(
M_{\gamma+\lambda}\right)  ^{2}}{n^{\gamma+2\lambda}}\right)  ^{-\frac{1}{2}}%
\end{align*}

On the other hand we have the following approximation for the integral%
\begin{align*}
&  \int_{\frac{L}{\left(  M_{\gamma+\lambda}\right)  ^{\frac{2}{\gamma
+2\lambda}}}}^{\frac{n}{\left(  M_{\gamma+\lambda}\right)  ^{\frac{2}%
{\gamma+2\lambda}}}}\log\left(  1+\frac{1}{y^{\gamma+2\lambda}}\right)  dy \\
%\\
%&  =\int_{0}^{\frac{n}{\left(  M_{\gamma+\lambda}\right)  ^{\frac{2}%
%{\gamma+2\lambda}}}}\log\left(  1+\frac{1}{y^{\gamma+2%\lambda}}\right)
%dy-\int_{0}^{\frac{L}{\left(  M_{\gamma+\lambda}\right)  %^{\frac{2}%
%{\gamma+2\lambda}}}}\log\left(  1+\frac{1}{y^{\gamma+2%\lambda}}\right)  dy\\
&  \simeq\int_{0}^{\frac{n}{\left(  M_{\gamma+\lambda}\right)  ^{\frac
{2}{\gamma+2\lambda}}}}\log\left(  1+\frac{1}{y^{\gamma+2\lambda}}\right)
dy+\left(  \gamma+2\lambda\right)  \int_{0}^{\frac{L}{\left(  M_{\gamma
+\lambda}\right)  ^{\frac{2}{\gamma+2\lambda}}}}\log\left(  y\right)  dy\\
%&  =\int_{0}^{\frac{n}{\left(  M_{\gamma+\lambda}\right)  ^{\frac{2}%
%{\gamma+2\lambda}}}}\log\left(  1+\frac{1}{y^{\gamma+2%\lambda}}\right)
%dy+\left(  \gamma+2\lambda\right)  \left[  y\log\left(  %y\right)  -y\right]
%_{0}^{\frac{L}{\left(  M_{\gamma+\lambda}\right)  ^{\frac{2}%{\gamma+2\lambda}%
%}}}\\
&  =\int_{0}^{\frac{n}{\left(  M_{\gamma+\lambda}\right)  ^{\frac{2}%
{\gamma+2\lambda}}}}\log\left(  1+\frac{1}{y^{\gamma+2\lambda}}\right)
dy+\frac{\left(  \gamma+2\lambda\right)  L}{\left(  M_{\gamma+\lambda}\right)
^{\frac{2}{\gamma+2\lambda}}}\log\left(  \frac{L}{\left(  M_{\gamma+\lambda
}\right)  ^{\frac{2}{\gamma+2\lambda}}}\right)  -\frac{\left(  \gamma
+2\lambda\right)  L}{\left(  M_{\gamma+\lambda}\right)  ^{\frac{2}%
{\gamma+2\lambda}}}%
\end{align*}
The error in the integral is bounded by%
\begin{align*}
&  \left(  M_{\gamma+\lambda}\right)  ^{\frac{2}{\gamma+2\lambda}}\int
_{0}^{\frac{L}{\left(  M_{\gamma+\lambda}\right)  ^{\frac{2}{\gamma+2\lambda}%
}}}y^{\gamma+2\lambda}dy 
%\\&  
%=C\left(  M_{\gamma+\lambda}\right)  ^{\frac{2}{\gamma+2%\lambda}}\left[
%\frac{L}{\left(  M_{\gamma+\lambda}\right)  ^{\frac{2}{\gamma+2\lambda}}%
%}\right]  ^{1+\gamma+2\lambda} %\\&  
=C\frac{L^{1+\gamma+2\lambda}}{\left(  M_{\gamma+\lambda}\right)  ^{2}} .%
\end{align*}
Therefore, this correction is negligible if we take $L$ such that 
$\frac{L^{1+\gamma+2\lambda}}{\left(  M_{\gamma+\lambda}\right)  ^{2}}\ll1.$

We now combine the previous approximation and we obtain
\begin{align*}
c_{n}  & \simeq 
%\simeq c_{L}\left(  \frac{L}{n}\right)  ^{\gamma+\lambda}%\exp\left(
%-\left(  M_{\gamma+\lambda}\right)  ^{\frac{2}{\gamma+2%%\lambda}}\int_{\frac
%{L}{\left(  M_{\gamma+\lambda}\right)  ^{\frac{2}{\gamma+2%\lambda}}}}%
%^{\frac{n}{\left(  M_{\gamma+\lambda}\right)  ^{\frac{2}%{\gamma+2\lambda}}}%
%}\log\left(  1+\frac{1}{y^{\gamma+2\lambda}}\right)  %dy\right) \times\\
%& \times\left(  1+\frac{\left(  M_{\gamma+\lambda}\right)  %^{2}}%
%{L^{\gamma+2\lambda}}\right)  ^{\frac{1}{2}}\left(  %1+\frac{\left(
%M_{\gamma+\lambda}\right)  ^{2}}{n^{\gamma+2\lambda}}\right) % ^{-\frac{1}{2}%
%}\\
%&  =
c_{L}\left(  \frac{L}{n}\right)  ^{\gamma+\lambda}\exp\left(  -\left(
M_{\gamma+\lambda}\right)  ^{\frac{2}{\gamma+2\lambda}}\int_{0}^{\frac
{n}{\left(  M_{\gamma+\lambda}\right)  ^{\frac{2}{\gamma+2\lambda}}}}%
\log\left(  1+\frac{1}{y^{\gamma+2\lambda}}\right)  dy\right) \times\\
& \quad\times\exp\left(  -\left(  \gamma+2\lambda\right)  L\log\left(  \frac
{L}{\left(  M_{\gamma+\lambda}\right)  ^{\frac{2}{\gamma+2\lambda}}}\right)
+\left(  \gamma+2\lambda\right)  L\right) \times\\
& \quad\times\left(  1+\frac{\left(  M_{\gamma+\lambda}\right)  ^{2}}%
{L^{\gamma+2\lambda}}\right)  ^{\frac{1}{2}}\left(  1+\frac{\left(
M_{\gamma+\lambda}\right)  ^{2}}{n^{\gamma+2\lambda}}\right)  ^{-\frac{1}{2}} \ .%
\end{align*}
On the other hand, using the formula above for $c_{L}$ and Stirling's formula we obtain%
\begin{align*}
c_{L} 
&  \simeq\frac{K}{L^{\gamma+\lambda}}\frac{\left(  L!\right)
^{\gamma+2\lambda}}{\left(  M_{\gamma+\lambda}\right)  ^{2L-1}} \\
& \simeq\frac
{K}{L^{\gamma+\lambda}}\left(  \sqrt{2\pi L}\right)  ^{\gamma+2\lambda}%
\exp\left(  \left(  \gamma+2\lambda\right)  L\log\left(  L\right)  -\left(
\gamma+2\lambda\right)  L-\left(  2L-1\right)  \log\left(  M_{\gamma+\lambda
}\right)  \right) \\
&  =\frac{\left(  2\pi\right)  ^{\frac{\gamma}{2}+\lambda}KM_{\gamma+\lambda
}\left(  L\right)  ^{\frac{\gamma}{2}+\lambda}}{L^{\gamma+\lambda}}\exp\left(
\left(  \gamma+2\lambda\right)  L\log\left(  L\right)  -2L\log\left(
M_{\gamma+\lambda}\right)  \right)  \exp\left(  -\left(  \gamma+2\lambda
\right)  L\right)
\end{align*}
Plugging this approximation above (and assuming that $L^{\gamma+2\lambda}%
\ll\left(  M_{\gamma+\lambda}\right)  ^{2}$, we obtain
\begin{comment}
\begin{align*}
c_{n}  &  \simeq\frac{\left(  2\pi\right)  ^{\frac{\gamma}{2}+\lambda
}KM_{\gamma+\lambda}\left(  L\right)  ^{\frac{\gamma}{2}+\lambda}}%
{L^{\gamma+\lambda}}\exp\left(  \left(  \gamma+2\lambda\right)  L\log\left(
L\right)  -2L\log\left(  M_{\gamma+\lambda}\right)  -\left(  \gamma
+2\lambda\right)  L\right) \times\\
& \times\left(  \frac{L}{n}\right)  ^{\gamma+\lambda}\exp\left(  -\left(
M_{\gamma+\lambda}\right)  ^{\frac{2}{\gamma+2\lambda}}\int_{0}^{\frac
{n}{\left(  M_{\gamma+\lambda}\right)  ^{\frac{2}{\gamma+2\lambda}}}}%
\log\left(  1+\frac{1}{y^{\gamma+2\lambda}}\right)  dy\right) \times\\
& \times\exp\left(  -\left(  \gamma+2\lambda\right)  L\log\left(  \frac
{L}{\left(  M_{\gamma+\lambda}\right)  ^{\frac{2}{\gamma+2\lambda}}}\right)
+\left(  \gamma+2\lambda\right)  L\right) \times\\
& \times\left(  1+\frac{\left(  M_{\gamma+\lambda}\right)  ^{2}}%
{L^{\gamma+2\lambda}}\right)  ^{\frac{1}{2}}\left(  1+\frac{\left(
M_{\gamma+\lambda}\right)  ^{2}}{n^{\gamma+2\lambda}}\right)  ^{-\frac{1}{2}}%
\end{align*}
or
\end{comment}
\begin{align*}
c_{n}  &  \simeq\frac{\left(  2\pi\right)  ^{\frac{\gamma}{2}+\lambda
}KM_{\gamma+\lambda}\left(  L\right)  ^{\frac{\gamma}{2}+\lambda}}%
{L^{\gamma+\lambda}}\exp\left(  L\log\left(  \frac{L^{\gamma+2\lambda}%
}{\left(  M_{\gamma+\lambda}\right)  ^{2}}\right)  \right) \times\exp\left(
-\left(  \gamma+2\lambda\right)  L\right) \\
& \times\left(  \frac{L}{n}\right)  ^{\gamma+\lambda}\exp\left(  -\left(
M_{\gamma+\lambda}\right)  ^{\frac{2}{\gamma+2\lambda}}\int_{0}^{\frac
{n}{\left(  M_{\gamma+\lambda}\right)  ^{\frac{2}{\gamma+2\lambda}}}}%
\log\left(  1+\frac{1}{y^{\gamma+2\lambda}}\right)  dy\right) \times\\
& \times\exp\left(  -L\log\left(  \frac{L^{\gamma+2\lambda}}{\left(
M_{\gamma+\lambda}\right)  ^{2}}\right)  +\left(  \gamma+2\lambda\right)
L\right) \times %\\&\times 
\left(  1+\frac{\left(  M_{\gamma+\lambda}\right)  ^{2}}%
{L^{\gamma+2\lambda}}\right)  ^{\frac{1}{2}}\left(  1+\frac{\left(
M_{\gamma+\lambda}\right)  ^{2}}{n^{\gamma+2\lambda}}\right)  ^{-\frac{1}{2}},
\end{align*}
and after some cancellations%
\begin{align*}
c_{n}  &  \simeq\frac{\left(  2\pi\right)  ^{\frac{\gamma}{2}+\lambda
}KM_{\gamma+\lambda}\left(  L\right)  ^{\frac{\gamma}{2}+\lambda}}%
{L^{\gamma+\lambda}}\left(  \frac{L}{n}\right)  ^{\gamma+\lambda}%
\exp\left(  -\left(  M_{\gamma+\lambda}\right)  ^{\frac{2}{\gamma+2\lambda}%
}\int_{0}^{\frac{n}{\left(  M_{\gamma+\lambda}\right)  ^{\frac{2}%
{\gamma+2\lambda}}}}\log\left(  1+\frac{1}{y^{\gamma+2\lambda}}\right)
dy\right) \times\\
& \times\left(  1+\frac{\left(  M_{\gamma+\lambda}\right)  ^{2}}%
{L^{\gamma+2\lambda}}\right)  ^{\frac{1}{2}}\left(  1+\frac{\left(
M_{\gamma+\lambda}\right)  ^{2}}{n^{\gamma+2\lambda}}\right)  ^{-\frac{1}{2}} \ . 
\end{align*}
Using that $\left(  L\right)  ^{\frac{\gamma}{2}+\lambda}\left(
1+\frac{\left(  M_{\gamma+\lambda}\right)  ^{2}}{L^{\gamma+2\lambda}}\right)
^{\frac{1}{2}}\simeq M_{\gamma+\lambda}$ for the range of values of $L$ under
consideration ($1\ll L^{\gamma+2\lambda}\ll\left(  M_{\gamma+\lambda}\right)
^{2}$), we obtain the approximation%
\begin{equation}
c_{n}\simeq\frac{\left[  \left(  2\pi\right)  ^{\frac{\gamma}{2}+\lambda
}K\right]  \left(  M_{\gamma+\lambda}\right)  ^{2}}{n^{\gamma+\lambda}%
\sqrt{1+\frac{\left(  M_{\gamma+\lambda}\right)  ^{2}}{n^{\gamma+2\lambda}}}%
}\exp\left(  -\left(  M_{\gamma+\lambda}\right)  ^{\frac{2}{\gamma+2\lambda}%
}\int_{0}^{\frac{n}{\left(  M_{\gamma+\lambda}\right)  ^{\frac{2}%
{\gamma+2\lambda}}}}\log\left(  1+\frac{1}{y^{\gamma+2\lambda}}\right)
dy\right)  \label{C4}%
\end{equation}
which gives the desired asymptotic formula for the stationary concentrations for $1\ll n$ (cf.\  \eqref{C7}).
%\bigskip
Notice that (\ref{C4}) implies the following asymptotics for $n\gg\left(
M_{\gamma+\lambda}\right)  ^{\frac{2}{\gamma+2\lambda}}$%
\begin{equation}
c_{n}\simeq\frac{\left[  \left(  2\pi\right)  ^{\frac{\gamma}{2}+\lambda
}K\right]  \left(  M_{\gamma+\lambda}\right)  ^{2}}{n^{\gamma+\lambda}}%
\exp\left(  -\left(  M_{\gamma+\lambda}\right)  ^{\frac{2}{\gamma+2\lambda}%
}\int_{0}^{\infty}\log\left(  1+\frac{1}{y^{\gamma+2\lambda}}\right)
dy\right).  \label{C5}%
\end{equation}

\bigskip 

\textbf{Declaration of interest} The authors declare that they have no conflict of interest.

\bigskip 

\textbf{Acknowledgements}
The authors gratefully acknowledge the support of the Hausdorff
Research Institute for Mathematics (Bonn), through the \textit{Junior Trimester Program on Kinetic Theory}, of the CRC 1060 \textit{The mathematics of emergent effects} at the University of
Bonn funded through the German Science Foundation (DFG), as well as of the \textit{Atmospheric
Mathematics} (AtMath) collaboration of the Faculty of Science of University of Helsinki. The
research has been supported by the Academy of Finland, via an Academy project (project
No. 339228) and the Finnish centre of excellence in \textit{Randomness and STructures} (project No.
346306). The research of M.A.F. has also been partially funded by the ERC Advanced Grant
741487 and the Centre for Mathematics of the University of Coimbra - UIDB/00324/2020
(funded by the Portuguese Government through FCT/MCTES). 
The funders had no role in study design, analysis, decision to publish,
or preparation of the manuscript.

\bibliographystyle{siam}

\bibliography{biblio}

 \bigskip

\textbf{M. A. Ferreira}  CMUC, Department of Mathematics, University of Coimbra,

\hspace{0.5 cm} 3000-413 Coimbra, Portugal

\hspace{0.5 cm} E-mail: marina.ferreira@mat.uc.pt

\hspace{0.5 cm} ORCID 0000-0001-5446-4845 

\hspace{0.5 cm} \textit{Previous affiliation}: Department of Mathematics and Statistics, University of Helsinki,

\hspace{0.5 cm} P.O. Box 68, FI-00014 Helsingin yliopisto, Helsinki, Finland

\bigskip 

\textbf{E. Franco}: Institute for Applied Mathematics, University of Bonn,

\hspace{0.5 cm} Endenicher Allee 60, D-53115 Bonn, Germany

\hspace{0.5 cm} E-mail: franco@iam.uni-bonn.de

\hspace{0.5 cm} ORCID 0000-0002-5311-2124

\bigskip 

\textbf{J. Lukkarinen}:  Department of Mathematics and Statistics, University of Helsinki,

\hspace{0.5 cm} P.O. Box 68, FI-00014 Helsingin yliopisto, Helsinki, Finland

\hspace{0.5 cm} E-mail: jani.lukkarinen@helsinki.fi

 \hspace{0.5 cm} ORCID 0000-0002-8757-1134

\bigskip 

\textbf{A. Nota}: Department of Information Engineering, Computer Science and Mathematics, 

\hspace{0.5 cm} University of L’Aquila, 67100 L’Aquila, Italy

\hspace{0.5 cm}  E-mail: alessia.nota@univaq.it

\hspace{0.5 cm}  ORCID 0000-0002-1259-4761

\bigskip 

\textbf{J. J. L. Vel\'azquez}: Institute for Applied Mathematics, University of Bonn,

\hspace{0.5 cm} Endenicher Allee 60, D-53115 Bonn, Germany

\hspace{0.5 cm} E-mail: velazquez@iam.uni-bonn.de

\end{document}